\documentclass[twocolumn,4pt,longbibliography,preprint,superscriptaddress,preprint,preprintnumbers,nobibnotes,amsmath,amssymb,aps, pra,tightenlines,nobalancelastpage]{revtex4-2}

\usepackage{graphicx}
\usepackage{dcolumn}
\usepackage{bm}

\usepackage{subfig}

\usepackage{bbold}

\usepackage{cancel}

\usepackage{hyperref}
\usepackage{xcolor}

\usepackage{enumerate}

\newcommand{\multiComment}[1]{}
\newcommand{\commutator}[2]{\left[ #1, #2\right]}
\newcommand{\Tr}[1]{\text{Tr}\left[ #1 \right]}
\newcommand{\Trp}[2]{\text{Tr}_{#1}\left[ #2 \right]}

\renewcommand{\det}[1]{\text{det}\left[ #1 \right]}

\newcommand{\QV}{Q^{\text{V}}}
\newcommand{\QAx}{Q^{\text{Ax}}}
\newcommand{\parity}{\mathbb{P}}
\newcommand{\Id}{\mathbb{1}}

\newcommand{\moment}[1]{\mathcal{Z}_{#1}}
\newcommand{\QAxMat}{\mathbb{Q}^{\text{Ax}}_A}
\newcommand{\QAxToeplitz}{T_{\ell}[\alpha]}
\newcommand{\QAxToeplitzMinus}{T_{\ell}[-\alpha]}
\newcommand{\Tau}{\mathcal{T}}

\newcommand{\ket}[1]{\left| #1 \right\rangle}

\newcommand{\GSM}[1]{\ket{\mathcal{M}=#1}}
\newcommand{\GSMGen}{\ket{\mathcal{M}}}
\newcommand{\GSD}{\ket{\text{GS}(\Delta)}}
\newcommand{\Z}[1]{\ket{Z=#1}}
\newcommand{\ZGen}{\ket{Z}}

\newcommand{\Pmat}{\mathcal{V}}

\begin{document}


\title{Entanglement asymmetry characterization of the chiral anomaly}

\author{Alfred Benedito}
\email{alfred.benedito@ift.csic.es}
\affiliation{\mbox{Instituto de F\'isica Te\'orica, UAM-CSIC, Universidad Aut\'onoma de Madrid, Spain.}}
\affiliation{\mbox{Quantum Advance Center Research, Madrid, Spain.}}
\author{Germ\'an  Sierra}%
\email{german.sierra@csic.es}
\affiliation{\mbox{Instituto de F\'isica Te\'orica, UAM-CSIC, Universidad Aut\'onoma de Madrid, Spain.}}
\affiliation{\mbox{Quantum Advance Center Research, Madrid, Spain.}}


\maketitle

\onecolumngrid 
\vspace{-3em}
\begin{center}
\begin{minipage}{0.9\linewidth}
\noindent
Shao et al. \cite{Shao_Original} recently showed that the 1+1D staggered fermion Hamiltonian admits a whole algebra of lattice operators that flow to the same axial charge in the thermodynamic limit (TL). On the lattice, the principal axial charge does not commute with the vector charge, although their commutator is expected to vanish in the TL, providing a lattice realization of the chiral anomaly. We investigate the effect of this anomaly on the ground-state sector using the entanglement asymmetry. Unexpectedly, the asymmetry remains nonzero in the TL, despite the vanishing of the commutator, and exhibits a novel scaling behavior.

\end{minipage}
\end{center}


\twocolumngrid 


\textit{Introduction.—} Quantum anomalies cannot be realized in strictly local lattice models with finite-dimensional local Hilbert spaces. This obstruction is readily illustrated by a $\mathrm{U}(1)$ symmetry in $(1+1)$ dimensions, for which the chiral anomaly appears as a Schwinger term $\commutator{j^0(t,x)}{j^1(t,x')}\sim i\partial_x\delta(x-x')$ \cite{Shao_Original,Shao_Original__ref9,Shao_Original__ref10}. In a finite-dimensional Hilbert space, however, the trace of any commutator vanishes, precluding such a central extension. A more rigorous formulation of this obstruction follows from the Nielsen-Ninomiya theorem \cite{Shao_Original,Shao_Original__ref12_a,Shao_Original__ref12_b,Shao_Original__ref13,Shao_Original__ref14}.

Nevertheless, the chiral anomaly leaves characteristic fingerprints on the lattice. It is encoded in an infinite hierarchy of increasingly nonlocal operators that generate the Onsager algebra and approach the axial (or vector) charge in the thermodynamic limit (TL). Each charge evades a distinct assumption of the Nielsen-Ninomiya theorem \cite{Shao_Original}. In particular, the main axial charge does so by failing to commute with the vector charge $\commutator{\QV}{\QAx}\ne 0$.

The anomaly thus admits two complementary characterizations: in the continuum, as the quantum breaking of a classical symmetry, and on the lattice, through the noncommutativity of the corresponding charges. This raises a natural question: can analogous signatures be detected using information-theoretic means? To address this question, we employ the (Rényi) entanglement asymmetry (EA) \cite{AsymOriginal_Ares2023}:
\begin{equation}
	\Delta S_n(\rho_A) = S_n(\rho_{A}^{\text{Sym}}) - S_n(\rho_A)
	\label{Rényi_Asym_Def}
\end{equation}
\noindent where $\rho_{A}^{\text{Sym}}$ is the symmetrized state (defined later) and $S_n(\rho)=\frac{1}{1-n}\log{\Tr{\rho^n}}$ is the $n$th Rényi entropy. The EA has been used to characterize symmetry breaking (SB) in critical theories \cite{EAsym_in_CFT_Fossati2024,EAsym_in_BCFT_Kusuki2025,EAsym_in_BCFT_Fossati2025,EAsym_in_BosonCFT_PhysRevD,Lastres_2025} and quantum circuits \cite{EAsym_RandQCircuits}, symmetry restoration following quantum quenches \cite{AsymOriginal_Ares2023,LackOfRestoration_SciPostPhys,XY_Mpemba_Murciano_2024,CyclicGroups_EAsym_Ferro_2024}, and the breaking of higher-form symmetries \cite{benini2026entanglementasymmetryhighernoninvertible,Casini2021,Stathis_propposed_n1,Stathis_propposed_n2}. It vanishes if and only if $\rho_A=\rho_A^{\mathrm{Sym}}$. Moreover, for MPS of finite bond dimension, the EA is determined solely by the symmetry-breaking pattern and is insensitive to other microscopic details \cite{Capizzi_2024}. 

\textit{Lattice model.—} We consider the microscopic Hamiltonian studied in \cite{Shao_Original}:
\begin{equation}
	H = -i\displaystyle\sum_{j=1}^N \left[ c_j^{\dagger}c_{j+1} + c_{j}c_{j+1}^{\dagger} \right]
	\label{crit_Hamiltonian}
\end{equation}
\noindent describing staggered spinless fermions on a chain with periodic boundary conditions (PBC). The model possesses a $\mathrm{U}(1)$ symmetry generated by the vector charge $\QV = \sum_{j=1}^N (c_j^{\dagger}c_j -1/2)$. This charge is \textit{site-local}, as its density is supported on individual lattice sites. For PBC, the model also admits an infinite hierarchy of conserved charges (in increasing order of non-locality) that flow towards the axial (or vector) charge in the TL. We focus on the most local representative:
\begin{align}
\QAx &= \frac{1}{2} \displaystyle\sum_{j=1}^N \left( c_j + c_j^{\dagger}\right) \left( c_{j+1} - c_{j+1}^{\dagger} \right) \nonumber \\
&= -\frac{i}{2} \displaystyle\sum_{j=1}^N a_{2j}a_{2j+1}
\label{Axial_Charge}
\end{align}
\noindent where the Majorana fermions $a_m = a_m^{\dagger}$ are defined by $c_j = 1/2 (a_{2j} -i a_{2j-1})$ and $\{a_m,a_n\} = 2\delta_{n,m}$. Both vector and axial charges are conserved $\commutator{H}{\QV} = \commutator{H}{\QAx} = 0$, whereas the anomaly is encoded in their non-commutativity $\commutator{\QV}{\QAx}\ne 0$. The axial charge is not \textit{site-local}, but it is \textit{link-local} (in the sense that its density is supported on individual lattice links).

\begin{figure}
	\centering
	\includegraphics[trim= 0.7cm 0.7cm 0.7cm 0.7cm, clip, scale=1.1]{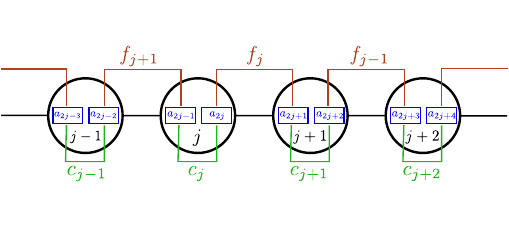}
	\caption{Majorana fermions $a_m$ associated with \textit{site} fermions $c_j$ and \textit{link} fermions $f_j$.}
	\label{Link_Fermion_Diagram}
\end{figure}

This \textit{link-local} structure is made explicit by the operator $T_b$ introduced in \cite{Shao_Original} such that $T_b a_{2j} T_b^{-1} = a_{2j},$ and $T_b a_{2j-1} T_b^{-1} = a_{2j+1}, \forall j$. It was originally interpreted as \textit{a right-moving charge conjugation times a continuum translation}: $T_b = \mathcal{C}^{R} \exp{(i2\pi \mathcal{P}/N)}$ \textit{on the low-lying states}, with the property:
\begin{equation}
	\QAx = T_b \QV T_b^{-1}
	\label{Tb_map_on_charges}
\end{equation}
We argue instead that, up to a global unitary, it is best understood as a \textit{half-step translation} corresponding to a Kramers-Wannier (KW) transformation \cite{KW_refs_1__RevModPhys,KW_refs_2__PhysRev,KW_refs_3__Fisher2004,HuaChen_paper} independently of any energy spectrum considerations. A proof based on a Jordan-Wigner mapping to a spin chain \cite{JW_paper} is presented in the Supplemental Material (SM). The same interpretation follows directly in the fermionic language: while $c_j$ creates a fermion associated with site $j$, the transformed operator $f_j:= T_b c_j T_b^{-1} = 1/2(a_{2j}-ia_{2j+1})$ creates a fermion associated with the \textit{link} between neighboring sites, as illustrated in Figure \ref{Link_Fermion_Diagram}. In terms of these \textit{link fermions}, the Hamiltonian retains its original form:
\begin{equation}
	H = -i\displaystyle\sum_{j=1}^N \left[ f_j^{\dagger}f_{j+1} + f_{j}f_{j+1}^{\dagger} \right]
	\label{crit_Hamiltonian_in_fs}
\end{equation}
\noindent which implies $\commutator{H}{T_b}=0$. Moreover, the vector and axial charges exchange their locality structures:
\begin{align}
\QV &= \frac{1}{2} \sum_{j=1}^N (f_j + f_j^{\dagger})(f_{j+1}-f_{j+1}^{\dagger})  \\
\QAx &= \sum_{j=1}^N \left( f_j^{\dagger} f_j -\frac{1}{2} \right)
\label{charges_in_f_language}
\end{align}

\textit{Axial charged moments: }Since the two charges do not commute, an eigenstate of $\QV$ is generally not an eigenstate of $\QAx$. Such a state therefore exhibits symmetry-resolved entanglement with respect to $\QV$ \cite{German_PhysRevB,GoldsteinSela_PhysRevLett,PhysRevD_categoricalSyms}, while having a nonvanishing EA with respect to $\QAx$. Within the ground-state (GS) subspace, however, the two charges commute: $P_{\text{GS}} \commutator{\QV}{\QAx} P_{\text{GS}} = 0$. Consequently, this subspace admits a common eigenbasis of $\QV$ and $\QAx$. These states provide natural candidates for probing anomalous subsystem properties through the EA, precisely because their axial EA vanishes at $\ell=N$ by construction. We therefore focus on $\ell<N$.

The link-local nature of $\QAx$ introduces an important distinction from an on-site symmetry. Consider a bipartition $(A,B)$ in which $A$ consists of the first $\ell$ sites. The axial charge supported strictly within $A$ and constructed from the same local density can only be $\QAx_A = \sum_{j=1}^{\ell-1}(f_j^{\dagger}f_j -1/2)$. The total axial charge does not decompose into independent contributions from $A$ and $B$. Instead, $\QAx = \QAx_A + \QAx_B + q_{\ell,\ell+1}^{\text{Ax}} + q_{N,1}^{\text{Ax}}$, where the last two terms cross the entanglement cuts. Accordingly, the corresponding symmetry transformation does not factorize $U^{\text{Ax}}\ne U^{\text{Ax}}_A \otimes U^{\text{Ax}}_B$. Consequently, even when the global state is symmetric, $\commutator{\QAx}{\rho}=0$, its reduced state need not be:
\begin{align}
\commutator{\rho_A}{\QAx_A} =& \Trp{B}{ \commutator{q_{\ell,\ell+1}^{\text{Ax}}}{\rho} + \commutator{q_{N,1}^{\text{Ax}}}{\rho} } \ne 0 
\label{failure_to_factorize}
\end{align}

Since $\QAx_A$ is well defined, one may symmetrize the reduced density matrix with respect to its eigenspaces:
\begin{equation}
	\rho_A^{\text{Sym}} := \displaystyle\sum_{q\in\sigma_A} P_q \rho_A P_q
\label{symmetrization_definition}
\end{equation}
\noindent where $\sigma_A$ denotes the spectrum of $\QAx_A$ and $P_q$ is the orthogonal projector onto the eigenspace with eigenvalue $q$. For charges whose eigenvalues are integers, the $P_q$ admit a Fourier representation. This is the case for $\QAx$:
\begin{equation}
	P_q = \displaystyle\int_{-\pi}^{\pi} \frac{d\alpha}{2\pi} e^{i \alpha (q - \QAx_A)}
\label{projectors}
\end{equation}
\noindent which allows calculation of the EA in terms of \textit{charged moments} $\moment{n}[\vec{\alpha}]$:
\begin{align}
\moment{n}[\vec{\alpha}] :=& \Tr{ \displaystyle\prod_{j=1}^{n} \rho_A e^{-i\alpha_{j,j+1} \QAx_A} } \label{chargedMoment} \\
\Tr{\left(\rho_A^{\text{Sym}}\right)^n} =& \displaystyle\int_{-\pi}^{\pi} \frac{d\alpha_1 \cdots d\alpha_n }{(2\pi)^n}\moment{n}[\vec{\alpha}]
\label{trace_with_moments}
\end{align}
\noindent where $\alpha_{i,j}=\alpha_i - \alpha_{j}$, $\vec{\alpha}=(\alpha_1,\ldots,\alpha_n)$ and $\alpha_{n+1}=\alpha_1$. The moments $\moment{n}$ can then be reduced to operations on the corresponding correlation matrices using standard techniques \cite{Fagotti_2010__CorrelationTech,AsymOriginal_Ares2023}. In other contexts, these moments are usually evaluated using Toeplitz-matrix methods, including the Widom--Szeg{\H{o}} limit theorems \cite{JinKorepin_2005,JinKorepin_ref37_p1_widom1974asymptotic,JinKorepin_ref37_p1_widom1974_ref_15_Szego,JinKorepin_ref37_p2_widom1974asymptotic,JinKorepin_ref38_boettcher}, the Fisher-Hartwig conjecture \cite{FisherHartwig_BASOR,FisherHartwig_OVCHINNIKOV,FilibertoPhD__FisherHartwig_PhysRevA} and generalizations \cite{Calabrese_2010,Calabrese_2010_ref_43__Generalized_FH,Bonsignori_2019,Bonsignori_2019_ref_27_basor}. For the EA, however, one generally encounters products of Toeplitz matrices, which are not themselves Toeplitz. The conjectures in \cite{LackOfRestoration_SciPostPhys} circumvent this issue by approximating products of Toeplitz matrices $T_{\ell}[\mathcal{G}_j]$ using products of their symbols $\mathcal{G}_j$:
\begin{align}
\log\det{ \Id + \displaystyle\prod_{j=1}^n T_{\ell}[\mathcal{G}_j] } \sim A \ell \label{Filiberto_logdet} \\
A = \displaystyle\int_{-\pi}^{\pi} \frac{dk}{2\pi} \log\det{ \Id + \displaystyle\prod_{j=1}^n \mathcal{G}_j(k) }  \label{Filiberto_A}
\end{align}

In the present setting, $\QAx_A$ is quadratic in the Majorana operators. Its exponential can therefore be written as $\exp{(-i\alpha \QAx_A)} = \exp{( a^T \alpha\QAxMat a /4)}$ where $\QAxMat$ is a $2\ell \times 2\ell$ matrix with elements $(\QAxMat)_{n,m}=\delta_{m,n-1}$ if $n\in\{3,5,...,2\ell-1\}$, $(\QAxMat)_{n,m}=-\delta_{m,n+1}$ if $n\in\{2,4,...,2\ell-2\}$ and $0$ everywhere else. The correlation matrix of the reduced state is $(\Gamma_A)_{n,m} := \Tr{\rho a_n a_m} - \delta_{n,m}$ with $n,m\in A$. Although $\Gamma_A$ is block Toeplitz in our case, $\QAxMat$ is not. In particular, the exponential $e^{\alpha\QAxMat}$ differs from a block-Toeplitz matrix at its two boundary blocks:
\begin{equation}
e^{\alpha \QAxMat} =
\begin{bmatrix}
    \check\Pi_{0}
    & \Pi_{-1}
    & 0
    & 0
    & \cdots
    & 0
    & 0
\\
    \Pi_{+1}
    & \Pi_{0}
    & \Pi_{-1}
    & 0
    & \cdots
    & 0
    & 0
\\
    0
    & \Pi_{+1}
    & \Pi_{0}
    & \Pi_{-1}
    & \cdots
    & 0
    & 0
\\
    \vdots
    & \vdots
    & \ddots
    & \ddots
    & \ddots
    & \vdots
    & \vdots
\\
    0
    & 0
    & 0
    & \cdots
    & \Pi_{+1}
    & \Pi_{0}
    & \Pi_{-1}
\\
    0
    & 0
    & 0
    & \cdots
    & 0
    & \Pi_{+1}
    & \hat\Pi_{0}
\end{bmatrix}
\label{exp_Axial}
\end{equation}
\noindent where the blocks are detailed in the SM. For $n=2$, \eqref{chargedMoment} simplifies to:
\begin{equation}
\moment{2}^2[\alpha] = \det{  \frac{\Id + \Gamma_A e^{\alpha \QAxMat} \Gamma_A e^{-\alpha \QAxMat }}{2} } 
\label{n2_SuperFormula}
\end{equation}
and the calculation would become analytically feasible if our matrices were Toeplitz. We devised a general method that allows for analytic calculations with matrices that are \textit{almost} (i.e., up to a finite set of blocks) block-Toeplitz by using Sylvester's theorem \cite{Sylvester_Thm}. We illustrate how it works in the SM. Applied to our case, it allows the second charged moment to be written exactly as
\begin{equation}
\moment{2}^2[\alpha] = \det{ \Tau_{\ell}[\alpha]/2 } \times o_{\ell}(\alpha)
\label{n2_SuperFormula_AfterSylvester}
\end{equation}
\noindent where the first factor is what $\moment{2}^2$ would be if the charge was fully block-Toeplitz and the second factor is the boundary correction. The matrix $\Tau_{\ell}[\alpha] := \left(\Id+\Gamma_A \QAxToeplitz \Gamma_A \QAxToeplitzMinus \right)$ contains the GS correlation matrix $\Gamma_A$ and $\QAxToeplitz$, obtained by replacing the boundary blocks in \eqref{exp_Axial} by $\Pi_0$. The factor $o_{\ell}(\alpha)$ is a determinant of a $4\times4$ matrix. Its explicit expression is given in the SM. Physically, $o_{\ell}(\alpha)$ captures the boundary contributions induced by the entanglement cuts.

\begin{figure}
	\centering
	\includegraphics[trim=0cm 0cm 0cm 0cm, clip, scale=0.22]{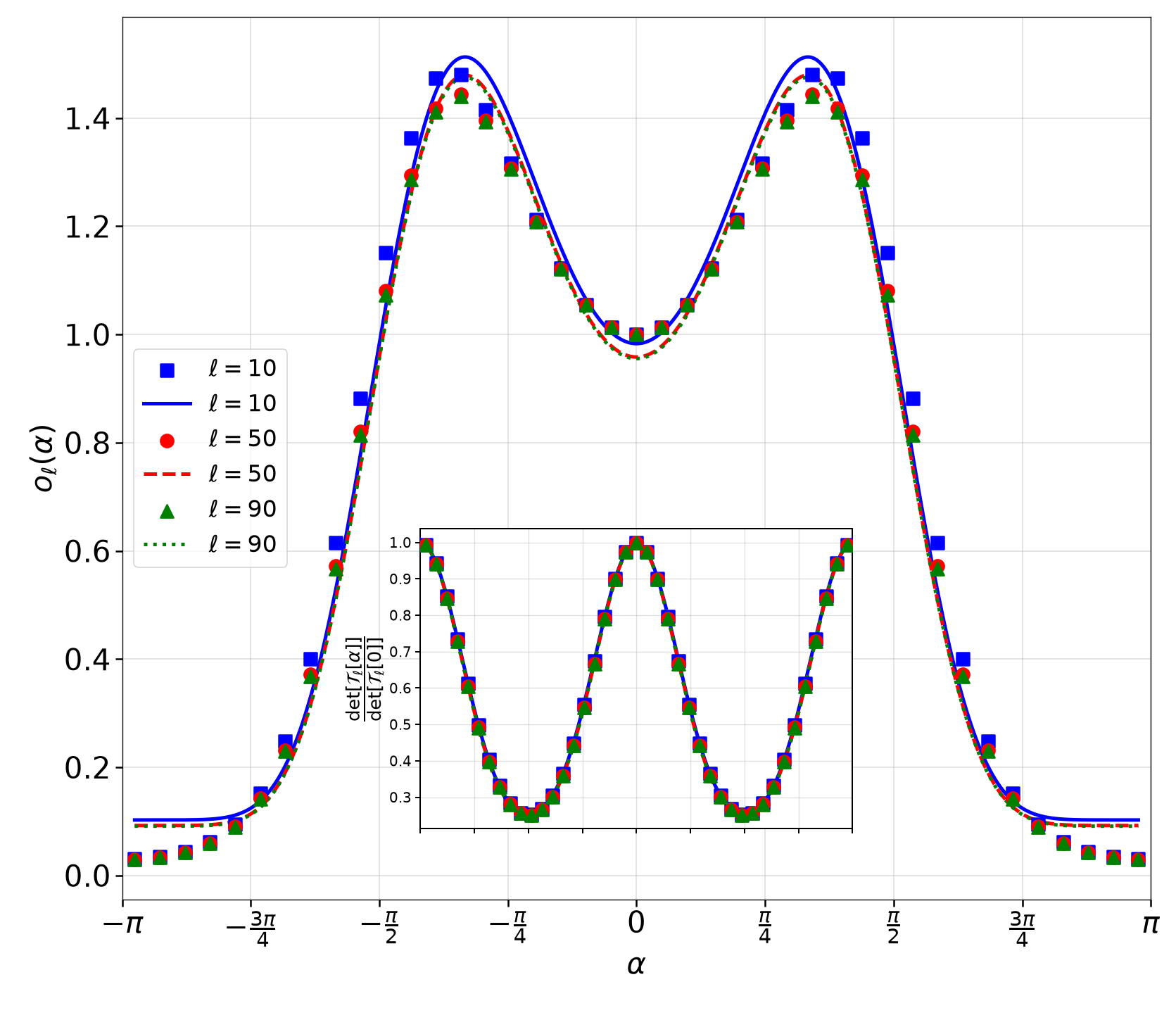}
	\caption{Numerical decomposition of $\moment{2}^{2}[\alpha]$ for the GS in the TL: $\det\Tau_{\ell}(\alpha)$ in the inset and $o_{\ell}(\alpha)$ in the main panel. The dependence on $\ell$ arises almost entirely from $o_{\ell}(\alpha)$ near the saddle points $\alpha=\pm\pi/3$. Symbols denote exact numerical results, while lines show the corresponding fits.}
	\label{ChargedMoments_Pieces_Fig}
\end{figure}

%
In the $\ell\to\infty$ limit, the conjectures in \cite{LackOfRestoration_SciPostPhys} apply to both determinants. For the GS in the TL (which is unique), $\Gamma_A$ and $\QAxToeplitz$ do not commute, whereas their Toeplitz symbols do. The conjecture (correctly) predicts a vanishing extensive rate \eqref{Filiberto_logdet}, $A=0$. For $o_{\ell}(\alpha)$, the conjectured symbol-product approximation yields (SM):
\begin{align}
\displaystyle\lim_{\ell\to\infty} o_{\ell}(\alpha) &= h^4(\alpha) \label{large_l_limit_o_is_h4} \\ 
h(\alpha) &= 1+\frac{\cos{\alpha} (1-\cos{\alpha})}{2} \label{large_l_limit_o}
\end{align}
As shown in Figure \ref{ChargedMoments_Pieces_Fig}, the $\Tau_{\ell}$ contribution is accurately described by $\left( \frac{\det{\Tau_{\ell}[\alpha]}}{\det{\Tau_{\ell}[0]}} \right) = B_{\ell} + A_{1,\ell}\cos^2{\alpha} + A_{2,\ell}\cos^4{\alpha}$ with fit coefficients $\{B_{\ell}, A_{p,\ell}\}_{p=1,2}$ nearly independent of $\ell$, while $o_{\ell}(\alpha) = C_{\ell} + Y_{\ell} h^4(\alpha)$ captures qualitatively the $\alpha$ dependence but not the $\ell$ dependence. Moreover, combining these results in a hybrid large-$\ell$ estimate does not reproduce the asymptotic $\beta$ constants in \eqref{even_Ninf_ansatz} and \eqref{odd_Ninf_ansatz}, obtained from the fits of the EA (SM). To the best of our knowledge, this is the first explicit example where the symbol-product approximation \cite{LackOfRestoration_SciPostPhys} is insufficient: the prediction \eqref{large_l_limit_o_is_h4} fails at $\mathcal{O}(1)$ when used to calculate $\beta_2$.
%
%
%

\begin{figure}
	\centering
	\includegraphics[trim=0cm 0cm 0cm 0cm, clip, scale=0.23]{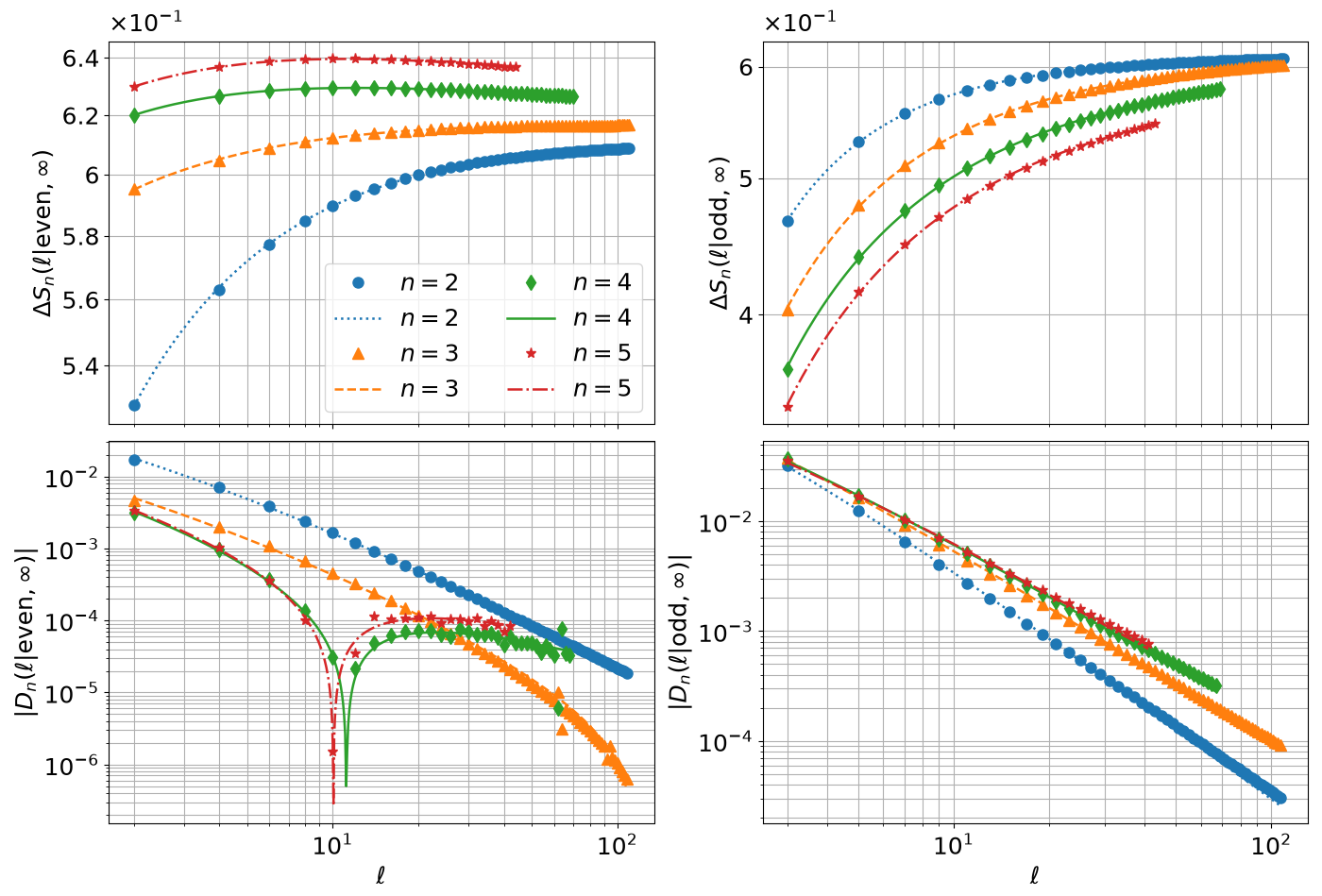}
	\caption{Axial EA of the GS in the TL. Symbols denote numerical data, while lines in the upper panels show fits to \eqref{even_Ninf_ansatz} and \eqref{odd_Ninf_ansatz}. The lower panels display the finite differences $D_n(\ell\mid\mathrm{p},\infty)$ obtained from the fitted curves. For $n=4,5$, $\ell_0\sim12,11$.}
	\label{Plot_TL_Asyms_fit}
\end{figure}

\textit{Axial EA in the TL.—} We evaluate the EA numerically, with implementation details provided in the SM. For the $n$th Rényi, the calculation generically involves $n$ nested weighted sums over the spectrum of $\QAx_A$. The resulting computational cost scales as $\mathcal{O}(|\sigma_A|^n)$. Nevertheless, we reach system sizes up to $N=110$ for $n=2,3$, $N=70$ for $n=4$, and $N=44$ for $n=5$, with corresponding maximal subsystem sizes in the TL calculations. The TL data are best described by the ansatz:
\begin{align}
\Delta S_n \left(\ell|\text{even},\infty \right) =& \frac{a_n \log{(\ell)} + \varepsilon_n}{\ell^{p_n}} + \beta_n \label{even_Ninf_ansatz} \\
\Delta S_n \left(\ell|\text{odd},\infty \right) =& \tilde{a}_n/\ell^{\tilde{p}_n} + \tilde{\beta}_n \label{odd_Ninf_ansatz} 
\end{align}
\noindent under the constraint that the functional form must be the same across $n$ for a given parity sector. Here $\beta_n \simeq \tilde{\beta}_n \simeq 0.61$ (see Table \ref{tab:deduced-beta-window} in the SM) for $n=2,3$. Finite differences $D_n(\ell,N) :=(\Delta S_n(\ell+2,N) - \Delta S_n(\ell,N))/2$ for $N\to\infty$ reveal the presence of the logarithmic correction in \eqref{even_Ninf_ansatz} through the stationary point $\ell_0(n)$ at which $|D_n(\ell=\ell_0(n),\infty)|=0$ (see Figure \ref{Plot_TL_Asyms_fit}).
Both parity sectors approach a nonzero EA as $\ell\rightarrow\infty$. At first sight, this may appear to conflict with the vanishing of the commutator between the vector and axial charges in the TL \cite{Shao_Original}. However, the original statement concerns the full system-size charges, while the EA probes the restriction to a subsystem. We interpret the asymptotic constant as a local remnant of the chiral anomaly in the lattice. Moreover, this entanglement cut causes the charge to be \textit{topological in the bulk but not in the boundary} (similarly to \cite{EAsym_in_BCFT_Kusuki2025}). Using the same picture of deforming insertions in the bulk to insertions of boundary-changing operators $\phi_b$, we conjecture that the $\tilde{p}_n$ in \eqref{odd_Ninf_ansatz} is related to the scaling dimension $\Delta_b$ of $\phi_b$, since these typically have contributions to the charged moments of the sort $\sim (\lambda/\ell)^{\Delta_b}$ (with $\lambda$ the UV cutoff).

\begin{figure}
	\centering
	\includegraphics[trim=0cm 0cm 0cm 0cm, clip, scale=0.3]{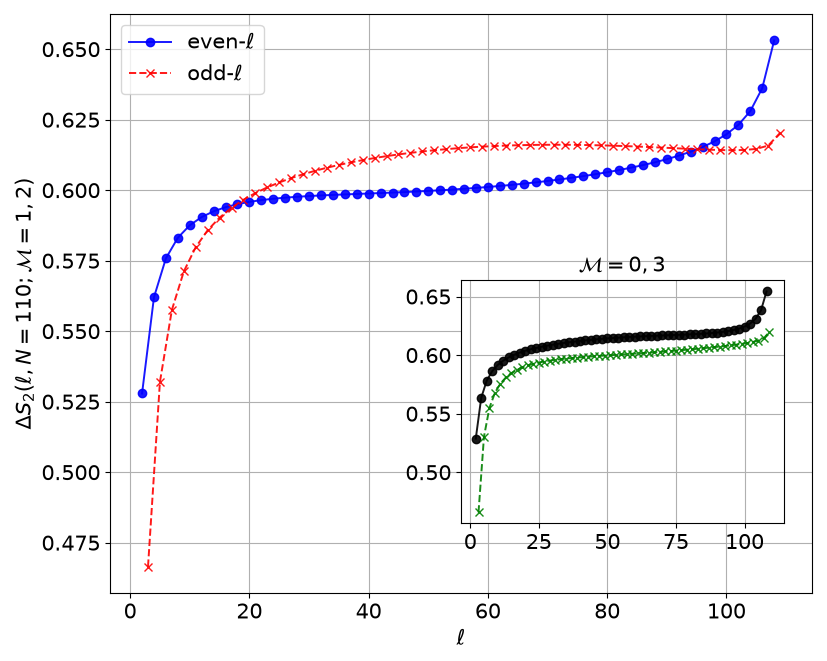}
	\caption{Axial EA of the GS for $N=110$ and $n=2$. For $\mathcal{M}=1,2$, the even- and odd-$\ell$ branches cross at $\ell/N\simeq1/4$ and $3/4$. No such crossings occur for $\mathcal{M}=0,3$ (inset).}
	\label{Plot_fN_Asyms_data_M_comparison}
\end{figure}

\textit{Axial EA at finite size.—} In contrast to the TL, the GS is degenerate at finite $N$. The GS degeneracy is fourfold for even $N$ and twofold for odd $N$. In the remainder of this Letter, we focus on even $N$. Degeneracy originates from zero modes at momenta $k=-\pi,0$. These GS are labeled as $\GSM{0}$ for the Fermi Sea (FS), $\GSM{1}=c_{-\pi}^{\dagger} \GSM{0}$, $\GSM{2}=c_{0}^{\dagger} \GSM{0}$ and $\GSM{3}=c_{-\pi}^{\dagger} c_{0}^{\dagger}\GSM{0}$. The EA is identical within the pairs $\mathcal{M}=0,3$ and $\mathcal{M}=1,2$. The latter pair exhibits a characteristic \textit{twist} relative to the former for all $n$. When separating even and odd parity curves, this manifests as the two curves crossing, as shown in Figure \ref{Plot_fN_Asyms_data_M_comparison}.
Focusing on $\GSM{0}$, we extract the finite-size corrections through $R_n(\ell,N):=\Delta S_n(\ell,N)-\Delta S_n(\ell,\infty)$. The corresponding raw data are presented in the SM. The data are well described by the scaling ansatz $R_n(\ell | \text{p} ,N) = B_{n,\text{p}}(N) \sin{(\pi x)} - A_{n,\text{p}}(N) \log{\frac{\sin{(\pi x)}}{\pi x}}$, where $x=\ell/(N-2)$ and p is even or odd. The amplitudes exhibit algebraic finite-size scaling $1/N^{q}$ with $q=q^{B/A}_{n,\text{p}}$ (see Figure \ref{two_panels_Fit_R_n_l_N__AmplitudeScaling} in the SM).

\textit{Axial EA mass-crossover.—} Finally, we explicitly break the chiral symmetry. In the Kogut--Susskind formulation, this is achieved by adding the staggered mass term \cite{Susskind_lattice_PhysRevD}:
\begin{equation}
V = \Delta \displaystyle\sum_{j=1}^N \left[ (-1)^j c_j^{\dagger} c_j \right], \quad \Delta\ge 0,
\label{staggeredMagnetization_massGap_term}
\end{equation}
\noindent to the Hamiltonian \eqref{crit_Hamiltonian}. The single-particle dispersion changes from $\varepsilon_k = 2\sin{(k)}$ to $E_k = \pm\sqrt{\varepsilon_k^2 + \Delta^2}$. $(H+V)$ can be diagonalized by a unitary transformation in $k$-space: $\left( c_k , c_{k+\pi} \right)^T \to \left( \alpha_k , \beta_k \right)^T$ for $k\in$ FS. For $\Delta>0$, $\GSD = \beta_0^{\dagger} \prod_{k\in\text{FS}} \beta_k^{\dagger} \ket{\Omega}$, satisfying $\GSD \to 1/\sqrt{2}\left( \GSM{1} - \GSM{2} \right)$ as $\Delta \to 0^+$. We follow instead the states $\{\ZGen\}$, defined such that $\ZGen \to \GSMGen$ as $\Delta \to 0^+$. More concretely, $\Z{0} := \beta_0 \GSD$, whose axial EA we study as a function of $\Delta$.
\begin{figure}
	\centering
	\includegraphics[trim=0cm 0cm 0cm 0cm, clip, scale=0.23]{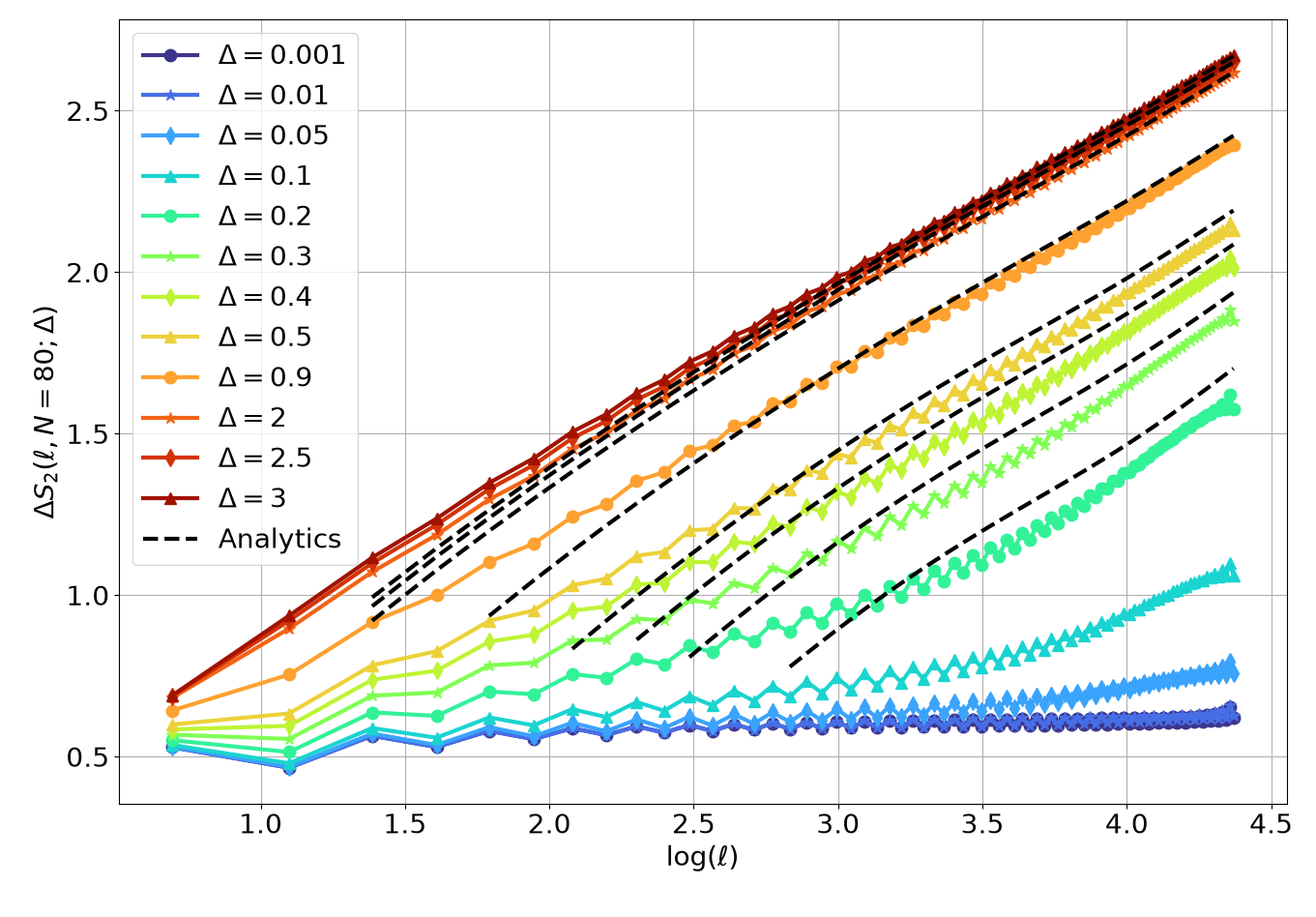}
	\caption{Axial EA of the state $\Z{0}$ at $\ell<N=80$ and $n=2$ for $\Delta\in[0.001,3]$. The analytical prediction is shown for $\Delta\ge 0.2$ for $\ell$ such that $p\ell>1.5$.}
	\label{Plot_fN_Asyms_data_Delta_comparison}
\end{figure}

When a $\mathrm{U}(1)$ symmetry is broken in the bulk, the EA scales as $\Delta S_n \simeq \frac{1}{2} \log{\ell}$ \cite{CyclicGroups_EAsym_Ferro_2024,AsymOriginal_Ares2023,Capizzi_2024,EAsym_in_CFT_Fossati2024}. By contrast, recent results for boundary symmetry breaking (BSB) of a $\mathrm{U}(1)$ symmetry \cite{EAsym_in_BCFT_Kusuki2025} yield $\Delta S_n \simeq \frac{1}{2}\log{\log{\ell}} + O(1)$. In that setting, the symmetry transformation factorizes across the bipartition, and the entanglement cut does not obstruct its topological character. For the axial charge considered here, by contrast, the entanglement cut itself generates a nonzero EA at criticality. We propose interpreting this behavior as \textit{anomalous boundary symmetry breaking} (ABSB), associated with the lattice realization of the chiral anomaly. Contrary to regular BSB, the scaling is not $\frac{1}{2}\log\log\ell$, but is instead given by \eqref{even_Ninf_ansatz} \& \eqref{odd_Ninf_ansatz}.

For large $\Delta$, one expects bulk SB scaling $\Delta S_n\simeq\frac{1}{2}\log\ell$. Both our analytical and numerical results exhibit this behavior when $\Delta>1$. By extending the method we introduced for non-block-Toeplitz corrections in the large $\ell$ limit (SM), we obtain for $\Z{0}$:
\begin{equation}
\Delta S_2 \simeq \begin{cases}
	\frac{1}{2}\log{\left( 2\pi p \ell \right)} - \frac{5-3p+8r_x}{8p\ell} & \text{if } \ell <N \\
	\frac{1}{2}\log{\left( \frac{\pi p N}{2} \right)} - \frac{9-3p}{8pN}  & \text{if } \ell =N \\
\end{cases}
\label{gapped_analytics}
\end{equation}
\noindent where $p = \Delta/(R_{\Delta} + \Delta)$, $R_{\Delta}=\sqrt{4+\Delta^2}$, $r_x = x^2/[(1-x)^2 + x^2]$ and $x=\ell/N$. This result comes from a saddle-point approximation (valid when $p\ell\gg 1$), so there are further contributions $\sim\mathcal{O}[(p\ell)^{-2}]$. For $0<\Delta\lesssim1$, the EA displays a smooth crossover between the critical and strongly gapped regimes, which we interpret as the axial symmetry going \textit{from} being topological in the bulk but obstructed at the entanglement cut \textit{to} not being topological anywhere. This is shown in Figure \ref{Plot_fN_Asyms_data_Delta_comparison} and captured by \eqref{gapped_analytics} for $\Delta\ge 0.2$. The EA of the GS can also be computed (SM) and, for $\ell=N$, it differs from the one reported in \cite{Sara_Chiral}. This is because our system maps to a sine-Gordon model \cite{SchwingerModel_bosonization_Coleman,SchwingerModel_bosonization_Coleman_Ref_1__sineGordon_original_equivalence} instead of a free massive boson.


\textit{Conclusions.—} In this Letter, we studied the chiral anomaly by means of the (Rényi) EA. We found the axial charge is naturally understood as the KW-dual of the vector charge at the critical point. The EA does not vanish in the TL, even though the commutator between vector and axial charges does. For finite-size systems, we characterized the scaling corrections and identified a characteristic \textit{twist} in the GS with a single zero mode occupied. We interpret these features of the critical behavior as \textit{anomalous boundary-symmetry-breaking}. Finally, upon explicitly breaking the chiral symmetry, we observed a smooth crossover from the critical behavior to the scaling $\frac{1}{2}\log(\ell)$ conventional for symmetry-broken bulk.

\textit{Acknowledgments:} We thank Filiberto Ares, Sara Murciano and Enrico Domanti for insightful conversations. We acknowledge financial support from the Spanish MINECO grant PID2021-127726NB-I00, PID2024-161474NB-I00 (MCIU/AEI/FEDER, UE), QUITEMAD-CM TEC2024/COM-84, the QUANTUM ENIA project Quantum Spain funded through the RTRP-Next Generation program under the framework of the Digital Spain 2026 Agenda and R\&D\&I Project CEX2025-001574-S, funded by MICIU/AEI/10.13039/501100011033. A.B. acknowledges support from the Spanish MICIU grant PRE2022-I0I93I. Computational resources were provided by the Drago cluster (SGAI-CSIC), the Hydra cluster at IFT, with support from Emilio Ambite, and Finisterrae III through CESGA.


\bibliographystyle{apsrev4-2}  
\bibliography{main}  

\newpage
\appendix

\onecolumngrid 
\newpage

\section{$T_b$ as a KW transformation and the \textit{link-fermion} picture}\label{section_Appendix_Kramers_Wannier}
We show that, up to a global unitary transformation, $T_b$ implements a Kramers--Wannier (KW) transformation \cite{KW_refs_1__RevModPhys,KW_refs_2__PhysRev,KW_refs_3__Fisher2004}. We first map the fermionic model to a spin-$1/2$ chain via a Jordan--Wigner (JW) transformation \cite{JW_paper}. Under this mapping, the Hamiltonian \eqref{crit_Hamiltonian} becomes a Dzyaloshinskii--Moriya interaction \cite{ToruMoriya_PhysRev,DZYALOSHINSKY_1958241}, with a boundary term controlled by the spin parity $\parity \equiv \mathbb{P}^z=\prod_{j=1}^N \sigma_j^z$. The vector charge $\QV$ maps to the magnetization, whereas $\QAx$ maps to a sum of $\sigma^y\sigma^y$ terms with the corresponding parity-dependent boundary contribution:

\begin{align}
H = -i\displaystyle\sum_{j=1}^N \left[ c_j^{\dagger}c_{j+1} + c_{j}c_{j+1}^{\dagger} \right] &\to \frac{1}{2} \displaystyle\sum_{j=1}^{N-1} \left( \sigma_j^x \sigma_{j+1}^y - \sigma_j^y \sigma_{j+1}^x \right) - \frac{1}{2} \parity \left( \sigma_N^x \sigma_{1}^y - \sigma_N^y \sigma_{1}^x \right)  \label{H_JW_map} \\
\QAx = -\frac{i}{2} \displaystyle\sum_{j=1}^N a_{2j}a_{2j+1} &\to  \frac{1}{2} \displaystyle\sum_{j=1}^{N-1} \left( \sigma_j^y \sigma_{j+1}^y \right) - \frac{1}{2} \parity \sigma_N^y \sigma_1^y \label{QAx_JW_map}\\
\QV = \sum_{j=1}^N (c_j^{\dagger}c_j -1/2) &\to  -\frac{1}{2} \displaystyle\sum_{j=1}^{N} \sigma_j^z \label{QV_JW_map}
\end{align}
\noindent We introduce the KW transformation, which maps spins on the sites $\sigma_j$ to spins on the links $S_{j,j+1}$. In the bulk:
\begin{align}
\sigma_j^y \sigma_{j+1}^y &= \tilde{S}_{j,j+1}^z = S_{j,j+1}^z \\
\sigma_j^z &= \tilde{S}_{j-1,j}^y \tilde{S}_{j,j+1}^y = S_{j-1,j}^x S_{j,j+1}^x
\label{KW_map}
\end{align}
\noindent where the $S$ operators are obtained from the $\widetilde{S}$ operators by a global unitary rotation about the $S^z$ axis. The transformation rules at the boundary impose on a periodic chain:
\begin{equation}
	\check\parity := \prod_{j=1}^N S_{j,j+1}^z = \prod_{j=1}^N \sigma_j^y \sigma_{j+1}^y = 1
	\label{restriction_1}
\end{equation}
\noindent i.e., a projection onto the even-parity sector of the Hilbert space.  Similarly, for the $x$-direction:
\begin{equation}
	\parity := \prod_{j=1}^N \sigma_j^z = \prod_{j=1}^N \left( S_{j-1,j}^x S_{j,j+1}^x \right)  = 1
	\label{restriction_2}
\end{equation}

The parity constraint renders the symmetry noninvertible. To characterize the action of $T_b$ in the spin representation, compare its fermionic action before and after the JW mapping:
\begin{align}
T_b\sigma_j^z T_b^{-1} &= -\sigma_j^y \sigma_{j+1}^y = -S_{j,j+1}^z \label{Tb_map_in_JW_spins_1}\\
T_b \sigma_j^y \sigma_{j+1}^y T_b^{-1} &= \sigma_j^y \sigma_{j+1}^z \sigma_{j+2}^y = -S_{j,j+1}^y S_{j+1,j+2}^y \label{Tb_map_in_JW_spins_3}
\end{align}
Equations \eqref{Tb_map_in_JW_spins_1} and \eqref{Tb_map_in_JW_spins_3} then give:
\begin{align}
\QV &= \displaystyle\sum_{j=1}^{N-1} \left( S_{j,j+1}^y S_{j+1,j+2}^y \right) - \frac{1}{2} \parity S_{N,1}^y S_{1,2}^y & \QAx &= T_b \QV T_b^{-1} = -\frac{1}{2} \displaystyle\sum_{j=1}^{N-1} S_{j,j+1}^z  \label{KW_check}
\end{align}
This is the spin counterpart of \eqref{charges_in_f_language}, establishing that $T_b$ implements the KW transformation.

\section{Finite-rank corrections to block-Toeplitz products}\label{section_Appendix_ToeplitzCalcs}
We develop a method for evaluating charged moments when one or more matrices entering a product differ from block-Toeplitz form by a finite-rank correction. The method separates the block-Toeplitz contribution, which can be treated using symbol-based asymptotics, from a finite-dimensional determinant encoding the non-Toeplitz part. We present the general setting, then illustrate directly how it works by applying it to the critical and gapped regimes discussed in the main text.\newline

\subsection{General finite-rank reduction}
Given an $m\times m$ matrix-valued symbol $\mathcal{G}(k)$, the corresponding $m\ell\times m\ell$ block-Toeplitz matrix is
\begin{equation}
T_{\ell}[\mathcal{G}] = \begin{bmatrix} 
    \Pi_0        & \Pi_{-1}     & \Pi_{-2}     & \cdots & \Pi_{-\ell+1} \\ 
    \Pi_{+1}     & \Pi_{0}      & \Pi_{-1}     & \cdots & \Pi_{-\ell+2} \\
    \vdots       & \vdots       & \vdots       & \ddots & \vdots  \\
    \Pi_{\ell-1} & \Pi_{\ell-2} & \Pi_{\ell-3} & \cdots & \Pi_{0}  \\
\end{bmatrix} \text{ where } \Pi_{n-m} = \displaystyle\int_{0}^{2\pi} \frac{dk}{2\pi} e^{-ik(n-m)} \mathcal{G}(k)
\label{generic_block_Toeplitz_matrix}
\end{equation}

We say a matrix $\tilde{T}_{\ell}[\mathcal{G}]$ is an \textit{almost-block-Toeplitz} matrix if it can be written as:
\begin{equation}
\tilde{T}_{\ell}[\mathcal{G}] = T_{\ell}[\mathcal{G}] + R_{\ell}
\end{equation}
\noindent where $T_{\ell}[\mathcal{G}]$ is a block-Toeplitz matrix and $R_{\ell}$ is a matrix with $\text{rank}(R_{\ell}) = \mathcal{O}(1)$ as $\ell$ grows. This should be contrasted with a generic $T_{\ell}[\mathcal{G}]$ whose rank is typically extensive:
\begin{equation}
	\text{rank}\left( T_{\ell}[\mathcal{G}] \right) = \mathcal{O}\left( \ell \times \text{rank}\left( \mathcal{G} \right) \right)
\end{equation}

Consider a product of block-Toeplitz matrices $\left\{ T_{\ell}[\mathcal{G}_j] \right\}_j$ in which an \textit{almost-block-Toeplitz} matrix is inserted at position $s$. Then:
\begin{align}
\det{\Id_{m\ell} + \left(\displaystyle\prod_{j=1}^{s-1} T_{\ell}[\mathcal{G}_j] \right) \tilde{T}_{\ell}[\mathcal{G}_s] \left(\displaystyle\prod_{j=s+1}^{n} T_{\ell}[\mathcal{G}_j] \right)} = \det{ \mathcal{M}_{\ell} } \times \nonumber \\
\times \det{ \Id_{m\ell} + \mathcal{M}_{\ell}^{-1} \left(\displaystyle\prod_{j=1}^{s-1} T_{\ell}[\mathcal{G}_j] \right) R_{\ell} \left(\displaystyle\prod_{j=s+1}^{n} T_{\ell}[\mathcal{G}_j] \right)}
\end{align}
\noindent provided that $\mathcal{M}_{\ell}$ is invertible, where 
\begin{equation}
\mathcal{M}_{\ell} := \Id_{m\ell} + \left(\displaystyle\prod_{j=1}^{n} T_{\ell}[\mathcal{G}_j] \right)
\end{equation}

The first factor $\det{ \mathcal{M}_{\ell} }$ contains pure block-Toeplitz contributions, while the second encodes the finite-rank corrections. The leading extensive contribution of the first factor can be estimated using the conjectures in \cite{LackOfRestoration_SciPostPhys}:
\begin{align}
\log\det{ \Id_{m\ell} + \displaystyle\prod_{j=1}^n T_{\ell}[\mathcal{G}_j] } &\sim A \ell,  & A &= \displaystyle\int_{-\pi}^{\pi} \frac{dk}{2\pi} \log\det{ \Id_{m} + \displaystyle\prod_{j=1}^n \mathcal{G}_j(k) },  \label{Filiberto_Appendix}
\end{align}

The finite-rank correction can be reduced to a determinant whose dimension does not scale with $\ell$. For  $\text{rank}(R_{\ell})=r$, we can write $R_{\ell} = \sum_{p=1}^r |e_p)(v_p|$ . Sylvester's theorem \cite{Sylvester_Thm} then gives that the finite-rank correction becomes $\det{\Id + K}$ where $K$ is an $r \times r$ matrix, with matrix elements:
\begin{equation}
K_{p,p'} = (v_p| \mathcal{Y}_{\ell} |e_{p'}) \qquad \mathcal{Y}_{\ell} := \left(\displaystyle\prod_{j=s+1}^{n} T_{\ell}[\mathcal{G}_j] \right) \mathcal{M}_{\ell}^{-1} \left(\displaystyle\prod_{j=1}^{s-1} T_{\ell}[\mathcal{G}_j] \right)
\end{equation}
\noindent which can be estimated within the same symbol-product approximation \cite{LackOfRestoration_SciPostPhys}:
\begin{equation}
K_{p,p'} \sim \displaystyle\sum_{x,y=1}^{\ell}\displaystyle\int_{-\pi}^{\pi} \frac{dq}{2\pi} e^{-iq(x-y)} (v_{p,m}(x)| \mathcal{G}_{\mathcal{Y}}(q) |e_{p',m}(y))
\end{equation}
\begin{equation}
\mathcal{G}_{\mathcal{Y}}(q) := \left(\displaystyle\prod_{j=s+1}^{n} \mathcal{G}_j \right) \left( \Id_{m} + \left(\displaystyle\prod_{j=1}^{n} \mathcal{G}_j \right)\right)^{-1} \left(\displaystyle\prod_{j=1}^{s-1} \mathcal{G}_j \right)
\end{equation}
\noindent where $x-y$ denotes an $m\times m$ block of the $\mathcal{Y}$ matrix and we decomposed the left and right vectors into their $m$-component block entries:
\begin{align}
(v_{p}| &= \bigoplus_{x=1}^{\ell} \text{ }_m (v_{p}(x)|, & |e_{p'}) &= \bigoplus_{y=1}^{\ell} |e_{p'}(y))_m
\end{align}

This general construction can be iterated for any finite number of \textit{almost-block-Toeplitz} insertions, provided that the required intermediate matrix inverses exist (although even this can be circumvented in certain cases, as we show when dealing with the gapped regime's $\Z{0}$ state). For $n$ insertions, $K$ is an $nr \times nr$ matrix instead. In the critical model considered below, the finite-rank corrections originate from the blocks adjacent to the entanglement cuts. More generally, the same reduction applies whenever block-translation invariance is violated only on a finite set of blocks, as occurs, for example, in the presence of a finite number of local defects or impurities.


\subsection{Critical-point boundary correction}
At the critical point, when $\ell=N$, there is no entanglement cut and the $\mathcal{M}$ states are eigenstates of $\QAx$, so the entanglement asymmetry vanishes for any system size $N$. We therefore restrict the remainder of this subsection to $\ell<N$ in the thermodynamic limit (TL) where the GS is unique. In that case, the correlation matrix (defined as $(\Gamma_A)_{n,m} := \Tr{\rho a_n a_m} - \delta_{n,m}$ where $n,m\in A$) is block-Toeplitz, with $2\times2$ blocks:
\begin{equation}
\Pi^{(GS)}_{n,m} = \frac{2 i \delta_{|n-m|,\text{odd}}}{\pi (n-m)} \Id_2; \to \mathcal{G}^{(GS)}(k) = f(k) \Id_2; \quad f(k) = \begin{Bmatrix}
	-1 & \text{if }0<k<\pi \\
	+1 & \text{if }\pi<k<2\pi \\
\end{Bmatrix};
\label{Toeplitz_state_info}
\end{equation}

The exponential of the subsystem axial charge has block structure:
\begin{equation}
e^{\alpha \QAxMat} =
\begin{bmatrix}
    \check\Pi_{0}
    & \Pi_{-1}
    & 0
    & 0
    & \cdots
    & 0
    & 0
\\
    \Pi_{+1}
    & \Pi_{0}
    & \Pi_{-1}
    & 0
    & \cdots
    & 0
    & 0
\\
    0
    & \Pi_{+1}
    & \Pi_{0}
    & \Pi_{-1}
    & \cdots
    & 0
    & 0
\\
    \vdots
    & \vdots
    & \ddots
    & \ddots
    & \ddots
    & \vdots
    & \vdots
\\
    0
    & 0
    & 0
    & \cdots
    & \Pi_{+1}
    & \Pi_{0}
    & \Pi_{-1}
\\
    0
    & 0
    & 0
    & \cdots
    & 0
    & \Pi_{+1}
    & \hat\Pi_{0}
\end{bmatrix}
\label{exp_Axial_appendix}
\end{equation}
\noindent where $\Pi_{0} = \cos{\alpha} \Id_2$, $\Id_2$ is the $2\times 2$ identity matrix and the other blocks are:
\begin{align}
	\check\Pi_{0} &=  \begin{pmatrix} 1 & 0 \\ 0 & \cos{\alpha} \end{pmatrix}; & \Pi_{-1} &= -\sin{\alpha} \begin{pmatrix} 0 & 0 \\ 1 & 0 \end{pmatrix}; \nonumber \\
	\hat\Pi_{0} &=  \begin{pmatrix} \cos{\alpha} & 0 \\ 0 & 1 \end{pmatrix}; & \Pi_{+1} &= \sin{\alpha} \begin{pmatrix} 0 & 1 \\ 0 & 0 \end{pmatrix};
\label{exp_QAx_blocks}
\end{align}

We decompose it as $\exp(\alpha \QAxMat) = T_{\ell}[\alpha] + R_{\ell}(\alpha)$. The block-Toeplitz contribution $T_{\ell}[\alpha]$ is obtained by replacing the two boundary blocks $\check\Pi_0$ and $\hat\Pi_0$ in \eqref{exp_Axial} with $\Pi_0$. The symbol of $T_{\ell}[\alpha]$ is:
\begin{equation}
\mathcal{G}_{\alpha}(k) = \begin{pmatrix} \cos{\alpha} & e^{+ik}\sin{\alpha} \\ -e^{-ik}\sin{\alpha} & \cos{\alpha} \end{pmatrix}
\label{T_l_axial}
\end{equation}

The remainder $R_{\ell}(\alpha)$ is a rank-two correction supported at the boundaries. Its only nonzero matrix elements are $(R_{\ell}(\alpha))_{n,m} = \delta_{\alpha}$ if $n=m=1$ or $n=m=2\ell$. Here $\delta_{\alpha} = 1-\cos{\alpha}$. Introducing the rank-one projectors $|1)(1|$ and $|2\ell)(2\ell|$ in correlation-matrix space, this can be written compactly as $R_{\ell}(\alpha) = \delta_{\alpha} \left[ |1)(1| + |2\ell)(2\ell| \right]$. We now apply this decomposition to the determinant entering the second charged moment:
\begin{equation}
\moment{2}^2[\alpha] = \det{\frac{\Id + \Gamma_A e^{\alpha \QAxMat}\Gamma_A e^{-\alpha \QAxMat}}{2}} = \det{\Tau_{\ell}[\alpha]/2} \times \det{\Id + \left( \Tau_{\ell}[\alpha] \right)^{-1}  \tilde{K}_{\ell}(\alpha) }
\end{equation}
\noindent valid as long as the $\Tau$-matrix is invertible. The $\Tau_{\ell}$ matrix is defined as:
\begin{align}
\Tau_{\ell}[\alpha] &:= \Id + \Gamma_A T_{\ell}[\alpha] \Gamma_A T_{\ell}[-\alpha] 
\end{align}
To reduce the boundary correction, define:
\begin{equation}
\Pmat := \begin{pmatrix} (1| \\ (2\ell| \end{pmatrix}; \quad \Pmat^{\dagger} = \begin{pmatrix} |1) & |2\ell) \end{pmatrix};
\end{equation}
\noindent of dimensions $2\times 2\ell$ and $2\ell \times 2$ respectively. Then:
\begin{equation}
\tilde{K}_{\ell}(\alpha) = \left( \Tau_{\ell}[\alpha] \right)^{-1} \times X Y^{\dagger}
\end{equation}
\noindent with
\begin{equation}
X = \left( \delta_{\alpha} \Gamma_A \Pmat^{\dagger}, \delta_{\alpha} \Gamma_A T_{\ell}[\alpha] \Gamma_A \Pmat^{\dagger} \right)
\label{X_definition}
\end{equation}
\noindent and
\begin{equation}
Y^{\dagger} = \begin{pmatrix} \Pmat \Gamma_A T_{\ell}[-\alpha] + \delta_{\alpha} \left( \Pmat \Gamma_A \Pmat^{\dagger} \right) \Pmat \\ \Pmat \end{pmatrix}
\label{Y_dagger_definition}
\end{equation}
Here, $X$ and $Y^{\dagger}$ have dimensions $2\ell \times 4$ and $4\times 2\ell$ respectively. Sylvester's theorem therefore yields:
\begin{equation}
o_{\ell}(\alpha) := \det{\Id_{2\ell \times 2\ell} + \tilde{K}_{\ell}(\alpha)} = \det{ \Id_{4\times 4}+ K_{\ell}(\alpha) }
\end{equation}
\noindent where $K_{\ell}(\alpha)$ is the $4\times 4$ matrix
\begin{equation}
K_{\ell}(\alpha) = Y^{\dagger} \left( \Tau_{\ell}[\alpha] \right)^{-1} X =: \textcolor{black}{\delta_{\alpha}} \textcolor{black}{K^{(1)}_{4\times 4}(\alpha)} + \textcolor{black}{\delta_{\alpha}}^2 \textcolor{black}{K^{(2)}_{4\times 4}(\alpha)} 
\label{starting_point_K_definition}
\end{equation}
Hence, the contribution generated by the entanglement boundaries is entirely encoded in $o_{\ell}$. All steps up to this point are exact. We now invoke the symbol-product approximation. In the large-$\ell$ limit, the symbol associated with $\mathcal{M}_{\ell}[\alpha]=\Tau_{\ell}[\alpha]/2$ is
\begin{equation}
\mathcal{G}_{\mathcal{M},\alpha}(k) \sim \frac{1}{2}\left[ \Id_{2\times 2} + \mathcal{G}^{(GS)}(k) \mathcal{G}_{\alpha}(k) \mathcal{G}^{(GS)}(k) \mathcal{G}_{-\alpha}(k) \right] = \Id_{2\times 2}
\end{equation}
\noindent where we used that $f^2(k) = 1$, $\forall k$ and $\mathcal{G}_{-\alpha}(k) = \mathcal{G}_{+\alpha}(k)^{-1}$. Therefore, \eqref{Filiberto_Appendix} predicts $A=0$ for $\det{\mathcal{M}_{\ell}[\alpha]}$'s extensive contribution.\newline

It remains to evaluate the matrix elements of $K_\ell(\alpha)$ using the symbol-product approximation. As an example, consider its upper-left $2\times2$ block:
\begin{equation}
\left[ K_{\ell}(\alpha) \right]_{1,1} = \delta_{\alpha} \Pmat \mathcal{A}^{(1)}_{\ell} \Pmat^{\dagger} + \delta_{\alpha}^2 G_{\partial A} \Pmat \mathcal{A}^{(2)}_{\ell} \Pmat^{\dagger}
\end{equation}
\noindent where $G_{\partial A} = \Pmat \Gamma_A \Pmat^{\dagger}$. At the symbol-product level:
\begin{align}
\left[\mathcal{A}^{(1)}\right]_{r} &\sim \displaystyle\int_0^{2\pi} \frac{dk}{2\pi} e^{-ikr} \left[ \mathcal{G}^{(GS)}(k) \mathcal{G}_{-\alpha}(k) \mathcal{G}_{\Tau,\alpha}(k)^{-1} \mathcal{G}^{(GS)}(k) \right]
\end{align}
\noindent and equivalently for $\mathcal{A}^{(2)}$. Projecting onto the boundaries:
\begin{align}
\left[ K^{(1)}_{4\times 4}(\alpha) \right]_{1,1} &= \begin{pmatrix} \mathcal{A}^{(1)}_{1,1} & \mathcal{A}^{(1)}_{1,2\ell} \\ \mathcal{A}^{(1)}_{2\ell,1} & \mathcal{A}^{(1)}_{2\ell,2\ell} \end{pmatrix} \sim \frac{1}{2}\cos{(\alpha)} \Id_{2\times 2} & \left[ K^{(2)}_{4\times 4}(\alpha) \right]_{1,1} & \sim  0_{2\times 2}
\end{align}

Repeating the same procedure for the remaining blocks yields:
\begin{equation}
\Id_{4\times 4} + K_{\ell}(\alpha) \simeq \begin{pmatrix} h(\alpha) \Id_{2\times 2} & \left( h(\alpha)-1 \right) G_{\partial A} \\ 0 & h(\alpha) \Id_{2\times 2} \end{pmatrix}
\end{equation}
\noindent whose determinant is
\begin{align}
o_\ell(\alpha) &\simeq h^4(\alpha) & h(\alpha) = \left[ 1 + \delta_{\alpha}\frac{\cos{\alpha}}{2} \right] \label{critical_o_GS}
\end{align}
This result captures the finite contribution generated by the entanglement boundaries but, as discussed in the main text, does not determine the subextensive $\ell$ dependence of $\det{\mathcal{M}_\ell[\alpha]}$. In particular, using only the boundary factor in the large $\ell$ limit gives
\begin{equation}
\Delta S_2 \simeq -\log\displaystyle\int_{-\pi}^{+\pi} \frac{d\alpha }{2\pi} h^2(\alpha) = +\log\frac{32}{23} \simeq 0.330
\end{equation}
\noindent which does not quantitatively reproduce the asymptotic constants obtained from a scaling analysis of the EA: $\beta_2 \simeq 0.61042 \pm 0.00008$ (even $\ell$) and $\tilde{\beta}_2 \simeq 0.610805\pm0.000016$ (odd $\ell$) (see Table \ref{tab:deduced-beta-window}). An improved estimate can be obtained by constructing a large-$\ell$ hybrid description for the charged moment combining $o_\ell(\alpha)$ with the numerical fit $\left( \frac{\det{\mathcal{M}_{\ell}[\alpha]}}{\det{\mathcal{M}_{\ell}[0]}} \right) = B_{\ell} + A_{1,\ell}\cos^2{\alpha} + A_{2,\ell}\cos^4{\alpha}$. The (almost) $\ell$-independent coefficients $\{ B_{\ell}, A_{1,\ell}, A_{2,\ell} \}$ saturate well before the largest accessible subsystem size is reached, giving:
\begin{equation}
\Delta S_2 \simeq -\log\displaystyle\int_{-\pi}^{+\pi} \frac{d\alpha }{2\pi} h^2(\alpha)\sqrt{B + A_1 \cos^2{\alpha} + A_2 \cos^4{\alpha}} 
\end{equation}
Using the coefficients obtained at $\ell=90$ (see Table \ref{tab:B-A1-A2-fit}), we estimate:
\begin{equation}
I \simeq 0.5006 \pm 0.0008 \implies \Delta S_2 \simeq 0.6919 \pm 0.0016
\end{equation}
Although this still differs quantitatively from the asymptotic constants in Table \ref{tab:deduced-beta-window}, it provides a substantially improved estimate and confirms that the boundary factor alone does not account for the full $\mathcal{O}(1)$ contribution.

\begin{table}
\centering
\begin{tabular}{c  c  c}
\hline
$B$ & $A_1$ & $A_2$ \\
\hline
$0.2527 \pm 0.0007$
&
$0.372 \pm 0.003$
&
$0.373 \pm 0.003$
\\
\hline
\end{tabular}
\caption{Fitted values of $B$, $A_1$, and $A_2$ for $\ell=90$.}
\label{tab:B-A1-A2-fit}
\end{table}

\subsection{The massive regime $\Delta>0$}
For $\Delta>0$, the staggered mass term breaks both the chiral symmetry and translation invariance by one lattice site, while two-site translation invariance remains. It is therefore convenient to group the lattice into two-site unit cells, so that the corresponding Majorana correlation matrices retain a block-Toeplitz structure with $4\times4$ blocks.

We consider the states
\begin{align}
\GSD &= \beta_0^{\dagger} \prod_{k\in\text{FS}} \beta_k^{\dagger} \ket{\Omega}, & \Z{0} &= \beta_0 \GSD,
\end{align}
\noindent where the $\beta_k$ come from one of the two quasiparticle branches. For the GS:
\begin{equation}
\Gamma^{(GS)}_{\text{cells}(x,y)} = \displaystyle\int_{-\pi}^{+\pi} \frac{dq}{2\pi} e^{-iq(x-y)} \mathcal{G}^{(GS)}(q), \qquad \mathcal{G}^{(GS)}(q) = \begin{bmatrix}
g(q) \sigma^y & f(q) \Id_2 \\
f(q)^* \Id_2 & -g(q)\sigma^y 
\end{bmatrix},
\label{gapped_Gamma_GS}
\end{equation}
\noindent where:
\begin{align}
g(q) &= \frac{\Delta}{E_{q/2}}, & f(q) &= -\frac{\varepsilon_{q/2} e^{iq/2}}{E_{q/2}}, &  E_k &= \sqrt{\Delta^2 + 4\sin^2(k)}, & \varepsilon_k &= 2\sin(k),
\end{align}
The $Z=0$ correlation matrix differs only by a small correction:
\begin{equation}
\Gamma^{(Z)}_{\text{cells}(x,y)} = \Gamma^{(GS)}_{\text{cells}(x,y)} + \delta\Gamma^{(Z)}_{\text{cells}(x,y)},\quad \delta\Gamma^{(Z)}_{\text{cells}(x,y)} = C_N  \begin{bmatrix}
\sigma^y & 0 \\
0 & 0 
\end{bmatrix}, \quad C_N = \frac{-4}{N}
\label{gapped_Gamma_Z}
\end{equation}
\begin{equation}
\Gamma^{(Z)}_{\text{cells}(x,y)} = \displaystyle\int_{-\pi}^{+\pi} \frac{dq}{2\pi} e^{-iq(x-y)} \mathcal{G}^{(Z)}(q), \qquad \mathcal{G}^{(Z)}(q) = \mathcal{G}^{(GS)}(q) + \delta\mathcal{G}^{(Z)}(q),
\label{gapped_Gamma_Z_symbol_part1}
\end{equation}
\begin{equation}
\delta\mathcal{G}^{(Z)}(q) = C_N (2\pi)\delta(q) \begin{bmatrix}
\sigma^y & 0 \\
0 & 0 
\end{bmatrix},
\label{gapped_Gamma_Z_symbol_part2}
\end{equation}

To match the two-site unit-cell representation introduced above, we regroup the $2\times2$ blocks of $T_\ell[\alpha]$ in \eqref{T_l_axial} into $4\times4$ blocks. The new matrix-symbol is
\begin{equation}
\tilde{\mathcal{G}}_{\alpha}(q) = \begin{bmatrix}
c_{\alpha}          & 0          & 0           & s_{\alpha} e^{iq} \\
0                   & c_{\alpha} & -s_{\alpha} & 0 \\
0                   & s_{\alpha} &  c_{\alpha} & 0 \\
-s_{\alpha} e^{-iq} & 0          & 0           & c_{\alpha} 
\end{bmatrix} 
\end{equation}
\noindent where $c_{\alpha} \equiv \cos{\alpha}$ and $s_{\alpha} \equiv \sin{\alpha}$.\newline

We now consider the second charged moment of the $\Z{0}$ state:
\begin{equation}
\moment{2}^2[\alpha,Z]=\det{\frac{\Id + \Gamma_A^{(Z)} e^{\alpha \QAxMat}\Gamma_A^{(Z)} e^{-\alpha \QAxMat}}{2}}  
\end{equation}
The $\Z{0}$ matrix-symbol contains a singular Dirac delta term \eqref{gapped_Gamma_Z_symbol_part2} which precludes a direct symbol-level inversion. Instead of using the prior procedure, we isolate the ground-state contribution and treat $\delta\Gamma_A^{(Z)}$ as a finite-rank correction. This gives the exact factorization
\begin{equation}
\moment{2}^2[\alpha,Z]=\moment{2}^2[\alpha,GS] \times \delta\moment{2}^2[\alpha,Z,GS]
\label{factorization_super_equation_1}
\end{equation}
\noindent with
\begin{equation}
\moment{2}^2[\alpha,GS]= 2^{-2\ell} \det{ \mathcal{M}_{\ell}(\alpha) }
\label{gapped_moment_GS}
\end{equation}
\begin{equation}
\delta\moment{2}^2[\alpha,Z,GS] = \det{ \Id + \left( \mathcal{M}_{\ell}(\alpha) \right)^{-1} W[\alpha]}
\label{gapped_moment_Z_perturbation}
\end{equation}
\noindent where the matrices are defined as:
\begin{align}
\mathcal{M}_{\ell}(\alpha) &:= \Id + \Gamma_A^{(GS)} e^{\alpha \QAxMat}\Gamma_A^{(GS)} e^{-\alpha \QAxMat} \label{M_GS_definition}
\end{align}
\begin{equation}
W[\alpha] := \left[ \delta\Gamma_A^{(Z)} e^{\alpha \QAxMat}\Gamma_A^{(Z)} e^{-\alpha \QAxMat} + \Gamma_A^{(GS)} e^{\alpha \QAxMat}\delta\Gamma_A^{(Z)} e^{-\alpha \QAxMat} \right]
\label{W_definition}
\end{equation}

Although the matrix elements of $\delta\Gamma_A^{(Z)}$ scale as $1/N$, the correction is coherent across the subsystem. Its contribution therefore need not vanish when $\ell$ scales with $N$; indeed, the relevant finite-size dependence will be controlled by the ratio $\ell/N$. The factorization in Eq.~\eqref{factorization_super_equation_1} isolates this contribution without introducing the singular $\delta(q)$ term into the Toeplitz-symbol inversion.

\subsubsection{Ground-state entanglement asymmetry}
We first evaluate the GS charged moment. Because the $\exp{(\alpha \QAxMat )}$ matrix is still not block-Toeplitz, we use the same decomposition into a pure block-Toeplitz matrix $\Tau_{\ell}$ and a finite-rank correction:
\begin{equation}
\moment{2}^2[\alpha,GS]=2^{-2\ell} \det{ \mathcal{M}_{\ell}(\alpha) }=2^{-2\ell} \det{ \Tau_{\ell}(\alpha) } o_{\ell}[\alpha,\Delta]
\label{gapped_moment_GS_boundary_factorization}
\end{equation}

The bulk contribution is:
\begin{equation}
\log\det{ \Tau_{\ell}(\alpha) } \sim A_{\alpha}[\Delta] \times \ell \quad \text{with} \quad A_{\alpha}[\Delta] = \displaystyle\int_{-\pi}^{\pi} \frac{dq}{2\pi} \log\det{t_{\alpha}(q,\Delta)}
\label{gapped_GS__bulk_A}
\end{equation}
\noindent with matrix-symbol $t_{\alpha}(q,\Delta)$
\begin{equation}
t_{\alpha}(q,\Delta) := \left[\Id_{4} + \mathcal{G}^{(GS)}(q) \tilde{\mathcal{G}}_{+\alpha}(q) \mathcal{G}^{(GS)}(q) \tilde{\mathcal{G}}_{-\alpha}(q)\right]
\end{equation}
\noindent and determinant
\begin{equation}
\det{t_{\alpha}(q,\Delta)} = 16\left[ 1- s_{\alpha}^2 g^2(q) \cos^2{(q/2)} \right]^2
\label{gapped_t_determinant}
\end{equation}

The momentum integral \eqref{gapped_GS__bulk_A} can be evaluated exactly:
\begin{equation}
A_{\alpha}[\Delta] = 4\log{(2)} + 4\log{\left[\mathcal{P}_{\Delta}(\alpha)\right]}, \quad \mathcal{P}_{\Delta}(\alpha) := \left( \frac{R_{\Delta} + \Delta |c_{\alpha}|}{R_{\Delta} + \Delta} \right), \quad R_{\Delta} = \sqrt{\Delta^2 + 4},
\label{gapped_bulk_GS_A_solved}
\end{equation}

Neglecting the boundary factor gives a bulk approximation for the asymmetry:
\begin{equation}
\Delta S_2^{\text{bulk}(GS)} \simeq -\log\displaystyle\int_{-\pi}^{\pi} \frac{d\alpha}{2\pi} \left[ \mathcal{P}_{\Delta}(\alpha) \right]^{\ell} \simeq \frac{1}{2}\log{\left(\frac{\pi}{2} p \ell \right)} - \frac{1-3p}{8p\ell} + \mathcal{O}\left( [p\ell]^{-2} \right)
\label{gapped_bulkOnly_EA_GS}
\end{equation}
\noindent where $p = \Delta/(R_{\Delta} + \Delta)$ and in the last step we performed a saddle-point approximation (valid when $p\ell\gg 1$) summing over the three existing saddle points $\alpha=0,\pm \pi$. For $\ell=N$, the bulk expression preceding the saddle-point approximation becomes exact, since no entanglement cuts are present.\newline

We now include the boundary correction. Applying Sylvester's theorem as in the critical case reduces $o_{\ell}[\alpha,\Delta]$ to a $4\times4$ determinant. The algebraic structure is identical to that of the critical case but with the critical symbols replaced by their massive counterparts. At large $\ell$, this gives a $K$-matrix \eqref{starting_point_K_definition}:
\begin{equation}
K^{(GS)}_{4\times 4}[\alpha,\Delta] \simeq \Id_2 \otimes \delta_{\alpha}  \begin{pmatrix}
\tilde{a}_{\Delta}(\alpha) & \tilde{b}_{\ell}(\alpha,\Delta) \\
-\tilde{b}_{\ell}(\alpha,\Delta) & \tilde{a}_{\Delta}(\alpha)
\end{pmatrix} 
\label{gapped_K_boundary_GS}
\end{equation}
\begin{align}
\tilde{a}_{\Delta}(\alpha) &:= \frac{\text{sign}(c_{\alpha})}{2} \left( \frac{R_{\Delta} |c_{\alpha}| + \Delta}{R_{\Delta}  + \Delta |c_{\alpha}| } \right) \label{gapped_tilde_a} \\
\tilde{b}_{\ell}(\alpha,\Delta) &:= \frac{s_{\alpha}\Delta}{2} \left(\frac{R_{\Delta} |c_{\alpha}| + \Delta}{\left[ R_{\Delta}  + \Delta |c_{\alpha}|  \right]^2} \right) \left[\rho_{\Delta}(\alpha)\right]^{\frac{\ell}{2}-1} & \rho_{\Delta}(\alpha) &= \left( \frac{R_{\Delta}  - \Delta |c_{\alpha}| }{R_{\Delta}  + \Delta |c_{\alpha}| } \right) \label{gapped_rho_for_tilde_b} 
\end{align}
\noindent where for $|c_{\alpha}|>0$, $0< \rho_{\Delta} <1$ meaning $\tilde{b}_{\ell}$ is exponentially suppressed with $\ell$. Therefore
\begin{equation}
o_{\ell}[\alpha,\Delta] = \det{\Id_4 + K^{(GS)}_{4\times 4}[\alpha,\Delta]} = \left[ \left( 1 + \delta_{\alpha} \tilde{a}_{\Delta}(\alpha) \right)^2 + \left( \delta_{\alpha} \tilde{b}_{\ell}(\alpha,\Delta) \right)^2 \right]^2
\label{gapped_o_GS}
\end{equation}
As a consistency check, the critical result \eqref{critical_o_GS} is recovered when $\Delta \to 0^+$
\begin{equation}
\lim_{\Delta\to 0^+} o_{\ell}[\alpha,\Delta] = h^4(\alpha) 
\end{equation}

Combining the bulk and boundary contributions gives the full ground-state asymmetry
\begin{equation}
\Delta S_2^{(GS)} = -\log\displaystyle\int_{-\pi}^{\pi} \frac{d\alpha}{2\pi} \left[ \mathcal{P}_{\Delta}(\alpha) \right]^{\ell} H_{\Delta}(\alpha)
\label{gapped_full_EA_GS_1}
\end{equation}
\noindent where we defined the boundary factor $H_{\Delta}(\alpha) := \sqrt{o_{\ell}[\alpha,\Delta]}$. At the saddle points $H_{\Delta}(\pm\pi) = 0$ and $H_{\Delta}(0) = 1$. The boundary factor therefore suppresses the saddles at $\alpha=\pm\pi$, leaving only the contribution around $\alpha=0$. Expanding both the bulk and boundary factors around $\alpha=0$ to compute the saddle-point approximation for the integral gives
\begin{equation}
\Delta S_2^{(GS)} \simeq \frac{1}{2}\log{\left( 2\pi p \ell \right)} - \frac{5-3p}{8p\ell} + \mathcal{O}\left( [p\ell]^{-2} \right) 
\label{gapped_full_EA_GS_2}
\end{equation}
Relative to the bulk approximation in \eqref{gapped_bulkOnly_EA_GS}, the boundary contribution changes both the constant inside the logarithm and the leading $1/\ell$ correction. The inverse symbol is singular at $q=0$, $\alpha=\pm\pi/2$. At finite boundary separation the corresponding Fourier coefficients are exponentially suppressed, $\sim\rho_{\Delta}(\alpha)^{\ell/2}$, so this singularity does not affect the saddle-point expansion about $\alpha=0$.\newline

Summarizing the results for the GS:
\begin{equation}
\Delta S_2^{(GS)} \simeq \begin{cases}
    \frac{1}{2}\log{\left( 2\pi p \ell \right)} - \frac{5-3p}{8p\ell} + \mathcal{O}\left( [p\ell]^{-2} \right) & \text{if } \ell <N \\
    \\
    \frac{1}{2}\log{\left(\frac{\pi}{2} p N \right)} - \frac{1-3p}{8pN} + \mathcal{O}\left( [pN]^{-2} \right) & \text{if } \ell =N 
\end{cases}
\label{gapped_full_EA_GS_2_allCases}
\end{equation}

\subsubsection{$\Z{0}$ entanglement asymmetry}
We now focus on calculating the correction $\delta\moment{2}$ required for the $\Z{0}$ state. The key simplification is that $\delta\Gamma_A^{(Z)}$ has rank 2. Indeed,
\begin{equation}
\frac{\delta\Gamma_A^{(Z)}}{C_N} = |\mathbf{1})(\mathbf{1}|_{\ell/2} \otimes \left[ \sigma^y \oplus 0_{2\times 2} \right] \equiv |+)(+| - |-)(-|
\end{equation}
\noindent where we defined the vectors (in correlation-matrix space):
\begin{align}
|\mathbf{1})_d &:= (1,1,\cdots,1)^T \in \mathbb{C}^{d\times 1} & |\pm)_4 &:= \frac{1}{\sqrt{2}}(1,\pm i,0,0)^T  & |\pm) &:= |\mathbf{1})_{\ell/2} \otimes |\pm)_4 
\end{align}

Using this projector decomposition on \eqref{W_definition}, $W$ can be written as:
\begin{equation}
W[\alpha] = C_N \mathbb{U} \mathbb{V}^{\dagger}
\end{equation}
\noindent where:
\begin{equation}
\mathbb{U} = \begin{pmatrix} |+),& \Gamma_A^{(GS)}e^{\alpha \QAxMat}|+),& -|-),& -\Gamma_A^{(GS)}e^{\alpha \QAxMat}|-) \end{pmatrix} \in \mathbb{C}^{2\ell \times 4}
\label{gapped_Z__U_definition}
\end{equation}
\begin{equation}
\mathbb{V}^{\dagger} = \begin{pmatrix}
(+| e^{\alpha \QAxMat} \Gamma_A^{(Z)} e^{-\alpha \QAxMat}  \\
(+| e^{-\alpha \QAxMat}   \\
(-| e^{\alpha \QAxMat} \Gamma_A^{(Z)} e^{-\alpha \QAxMat}  \\
(-| e^{-\alpha \QAxMat}   
\end{pmatrix} \in \mathbb{C}^{4\times 2\ell}
\label{gapped_Z__V_definition}
\end{equation}
Sylvester's theorem then reduces the correction to:
\begin{equation}
\delta\moment{2}^2[\alpha,Z,GS] = \det{ \Id + \left( \mathcal{M}_{\ell}(\alpha) \right)^{-1} W[\alpha]} = \det{\Id_4 + C_N K^{(Z)}_{4\times 4}}
\end{equation}
\noindent where
\begin{equation}
K^{(Z)}_{4\times 4} :=  \mathbb{V}^{\dagger} \left( \mathcal{M}_{\ell}(\alpha) \right)^{-1} \mathbb{U} 
\label{gapped__K_Z_definition}
\end{equation}
\noindent with matrix elements of the form
\begin{align}
\mathcal{A}_{\sigma,\tau} &:= (\sigma| e^{\alpha \QAxMat} \Gamma_A^{(Z)} e^{-\alpha \QAxMat} \left( \mathcal{M}_{\ell}(\alpha) \right)^{-1} |\tau)   \\
\mathcal{B}_{\sigma,\tau} &:= (\sigma| e^{\alpha \QAxMat} \Gamma_A^{(Z)} e^{-\alpha \QAxMat} \left( \mathcal{M}_{\ell}(\alpha) \right)^{-1} \Gamma_A^{(GS)} e^{\alpha \QAxMat} |\tau) \\
\mathcal{C}_{\sigma,\tau} &:= (\sigma| e^{-\alpha \QAxMat} \left( \mathcal{M}_{\ell}(\alpha) \right)^{-1} |\tau) \\
\mathcal{D}_{\sigma,\tau} &:= (\sigma| e^{-\alpha \QAxMat} \left( \mathcal{M}_{\ell}(\alpha) \right)^{-1} \Gamma_A^{(GS)} e^{\alpha \QAxMat} |\tau) 
\end{align}

Neither $\exp{(\pm\alpha\QAxMat)}$ nor $\mathcal{M}_{\ell}(\alpha)^{\pm 1}$ are block-Toeplitz. Nonetheless, we estimate the $\mathcal{F}_{\sigma,\tau} \in \left\{ \mathcal{A}_{\sigma,\tau}, \mathcal{B}_{\sigma,\tau}, \mathcal{C}_{\sigma,\tau}, \mathcal{D}_{\sigma,\tau} \right\}$ matrix elements by approximating any matrix by its block-Toeplitz part and subsequently applying the symbol-product approximation. This gives:
\begin{equation}
\mathcal{F}_{\sigma,\tau} = \displaystyle\int_{-\pi}^{+\pi} \frac{dq}{2\pi} \left[ \displaystyle\sum_{x,y=1}^{\ell/2} e^{-iq(x-y)} \right] \text{ }_4 (\sigma| \left[ \displaystyle\prod_j \mathcal{G}_j(q) \right] |\tau)_4 
\label{kernel_start}
\end{equation}
\noindent where, because the $|\pm)_4$ internal states are always the same, the $x,y$-sum factorizes into the Fejér kernel \cite{FejerKernel_ref__Katznelson_2004}:
\begin{equation}
\left[ \displaystyle\sum_{x,y=1}^{\ell/2} e^{-iq(x-y)} \right] = \frac{\ell}{2} F_{\ell/2}(q), \qquad F_{\mathcal{N}}(q) := \frac{1}{\mathcal{N}} \left[ \frac{\sin{\mathcal{N}q/2}}{\sin{q/2}} \right]^2
\label{Fejer_kernel}
\end{equation}
\noindent which has the property $\lim_{\mathcal{N}\to\infty} F_{\mathcal{N}}(q) = (2\pi)\delta(q)$. The leading contribution in the large $\ell$ limit becomes:
\begin{equation}
\mathcal{F}_{\sigma,\tau} \simeq \frac{\ell}{2} \text{ }_4 (\sigma| \left[ \displaystyle\prod_j \mathcal{G}_j(0) \right] |\tau)_4
\label{kernel_end}
\end{equation}
Thus the coherent support of the rank-two perturbation over the entire subsystem produces matrix elements that scale as $O(\ell)$. Keeping only this leading contribution yields
\begin{equation}
K^{(Z)}_{4\times 4} \simeq \begin{pmatrix}
\frac{\ell}{4} & 0 & \frac{\ell}{4c_\alpha} \left[ c_\alpha^2 \left( 2+\frac{C_N\ell}{2} \right)-1 \right] & 0 \\
0 & \frac{\ell}{4} & 0 & -\frac{\ell}{4c_\alpha} \left[ c_\alpha^2 \left( 2+\frac{C_N\ell}{2} \right)-1 \right] \\
\frac{\ell}{4c_\alpha} & 0 & \frac{\ell}{4} +\frac{C_N\ell^2}{8} & 0 \\
0 & -\frac{\ell}{4c_\alpha} & 0 & \frac{\ell}{4} +\frac{C_N\ell^2}{8}
\end{pmatrix}
\end{equation}
\begin{equation}
\delta\moment{2}^2[\alpha,Z,GS] = \det{\Id_4 + C_N K^{(Z)}_{4\times 4}} \simeq  \left[ \left(1-\frac{\ell}{N}\right)^2 + \frac{\ell^2}{N^2\cos^2\alpha} \right]^2
\label{gapped_Z_correction}
\end{equation}
The apparent singularity at $\alpha=\pm\pi/2$ is an artifact of replacing the Fejér kernel by its $\delta$-function limit before performing the momentum integral. Retaining the finite-width kernel regularizes this behavior. Since the asymptotic EA is controlled by the saddles located at $\alpha=0$ when $\ell<N$ and $\alpha=0,\pm\pi$ when $\ell=N$, this singularity does not affect the expansion below.\newline

The $\Z{0}$ correction will therefore appear in the $\alpha$-integral as a factor:
\begin{equation}
Q_{\Delta}[\alpha,x] = \frac{\delta\moment{2}[\alpha,Z,GS]}{\delta\moment{2}[0,Z,GS]} \simeq \frac{ \left(1-x\right)^2 + x^2/c_\alpha^2 }{ \left(1-x\right)^2 + x^2 }
\label{gapped_Z_Integral_Factor}
\end{equation}
\noindent where $x=\ell/N$. Finally, the full result is
\begin{align}
\Delta S_2^{(Z)} &= \begin{cases}
	-\log\displaystyle\int_{-\pi}^{\pi} \frac{d\alpha}{2\pi} \left[ \mathcal{P}_{\Delta}(\alpha) \right]^{\ell} H_{\Delta}(\alpha)Q_{\Delta}[\alpha,x<1] & \text{if } \ell <N \\
    \\
	-\log\displaystyle\int_{-\pi}^{\pi} \frac{d\alpha}{2\pi} \left[ \mathcal{P}_{\Delta}(\alpha) \right]^{N} Q_{\Delta}[\alpha,x=1] & \text{if } \ell =N \\
\end{cases} \nonumber \\
 & \simeq\begin{cases}
	\frac{1}{2}\log{\left( 2\pi p \ell \right)} - \frac{5-3p+8r_x}{8p\ell} + \mathcal{O}\left( [p\ell]^{-2} \right) & \text{if } \ell <N \\
    \\
	\frac{1}{2}\log{\left( \frac{\pi p N}{2} \right)} - \frac{9-3p}{8pN} + \mathcal{O}\left( [pN]^{-2} \right) & \text{if } \ell =N \\
\end{cases}
\end{align}
\noindent where $r_x = x^2/[(1-x)^2 + x^2]$.

\newpage

\section{Numerical details}\label{section_Appendix_NumericalDetails}
In this section, we describe the numerical evaluation of the charged moments and the fitting procedures used in the main text. The key insight is that, for symmetries whose eigenvalues are equally spaced integers, the projector onto a given eigenspace can be written either in the integral form of Eq.~\eqref{projectors} or as a discrete sum. This follows from the discrete orthogonality of roots of unity.
\begin{equation}
	P_q = \frac{1}{|\sigma_A|} \displaystyle\sum_{\alpha \in \frac{2\pi}{\mathcal{N}}\sigma_A} e^{i\alpha \left( q - \QAx_A \right)}
\label{discrete_rep_projector}
\end{equation}
\noindent where $\sigma_A$ denotes the spectrum of $\QAx_A$ and $\mathcal{N} = |\sigma_A|$ (i.e., $\mathcal{N}=\ell$ if $\ell<N$ and $\mathcal{N}=N+1$ if $\ell=N$). The corresponding discrete form of \eqref{trace_with_moments} is:
\begin{equation}
\Tr{\left(\rho_A^{\text{Sym}}\right)^n} = \frac{1}{\mathcal{N}^n} \displaystyle\sum_{q\in \sigma_A} \text{ } \displaystyle\sum_{\alpha_{1},\cdots,\alpha_{n}\in \frac{2\pi}{\mathcal{N}}\sigma_A} \text{ } e^{iq\Sigma[\vec{\alpha}]} \moment{n}[\vec{\alpha}] 
\label{discrete_charged_moments}
\end{equation}
\noindent where $\vec{\alpha}=(\alpha_{1},...,\alpha_{n})$ and $\Sigma[\vec{\alpha}] = \sum_{j=1}^n \alpha_{j}$. The number of $\vec{\alpha}$ configurations that must be summed over grows as $\mathcal{O}(\mathcal{N}^n)$. Efficiently distributing these configurations across parallel processes constitutes the principal computational bottleneck at large $n$.\newline

%
To determine the scaling form of the thermodynamic-limit data, we tested many physically motivated scaling forms, including algebraic, logarithmic, and double-logarithmic corrections as well as those same forms modulated by $1/\ell$, $1/\sqrt{\ell}$ and $1/\ell^{p_n}$. The derivative of the curves $d_n(\ell,N) :=d(\Delta S_n(\ell,N))/d\ell$ (which we approximate as $D_n(\ell,N) :=(\Delta S_n(\ell+2,N) - \Delta S_n(\ell,N))/2$ in the data) for $N\to\infty$ reveals the presence of the logarithmic correction in \eqref{even_Ninf_ansatz} through the stationary point $\ell_0(n)$ at which $|D_n(\ell=\ell_0(n),\infty)|=0$. By approximating $d_n(\ell,\infty)\simeq D_n(\ell,\infty)$, we obtain $\ell_0(n) \simeq \exp{\left( p_n^{-1} - \varepsilon_n/a_n \right)}$. The presence of this stationary point within the accessible data strongly stabilizes the fit. This occurs for $n=4,5$, for which $\ell_0\sim12,11$ respectively, as shown in Figure \ref{Plot_TL_Asyms_fit}. We required the functional form to be common across Rényi indices within each parity sector and to reproduce the finite differences $D_n(\ell,\infty)$ without additional fitting. Across all these ansätze, the asymptotic constant $\beta_n$ remained close to $0.6$. After this selection process, the functional form that best satisfied all these requirements was that of Eqs. \eqref{even_Ninf_ansatz} and \eqref{odd_Ninf_ansatz}. The parameters obtained by fitting the TL data are reported in Tables \ref{tab:even_l_TL_fit} and \ref{tab:odd_l_TL_fit}, respectively.\newline

\begin{table}[h]
\centering
\small
\begin{tabular}{c c c c c c c}
\hline
$n$
& $a_n$
& $\varepsilon_n$
& $\beta_n$
& $p_n$
& $\ell_0$
& $R^2$ \\
\hline

2
& $0.0316 \pm 0.0004$
& $-0.1405 \pm 0.0003$
& $0.60827 \pm 0.00013$
& $0.562050$
& $508.500^{\dagger}$
& $0.99982883$
\\

3
& $0.01315 \pm 0.00018$
& $-0.03595 \pm 0.00014$
& $0.61412 \pm 0.00009$
& $0.491830$
& $117.550^{\dagger}$
& $0.99928787$
\\

4
& $0.02249 \pm 0.00010$
& $-0.00676 \pm 0.00006$
& $0.61361 \pm 0.00007$
& $0.455635$
& $12.1275$
& $0.99942944$
\\

5
& $0.02883 \pm 0.00012$
& $0.01063 \pm 0.00009$
& $0.60605 \pm 0.00015$
& $0.361417$
& $11.0051$ 
& $0.99974743$
\\

\hline
\end{tabular}%

\caption{Fit parameters for the ansatz in  Eq. \eqref{even_Ninf_ansatz}. The reported uncertainties are conditional least-squares errors evaluated at fixed $p_n$. The uncertainties in $p_n$ were not estimated. A dagger $\dagger$ indicates that $\ell_0$ lies outside the accessible range.}
\label{tab:even_l_TL_fit}
\end{table}

\begin{table}
\centering
\small
\begin{tabular}{c c c c c}
\hline
$n$
& $\tilde{a}_n$
& $\tilde{\beta}_n$
& $\tilde{p}_n$ 
& $R^2$
\\
\hline

2
& $-0.498 \pm 0.001$
& $0.60973 \pm 0.00005$
& $1.13608$ 
& $0.99979747$
\\

3
& $-0.5161 \pm 0.0011$
& $0.61178 \pm 0.00010$
& $0.831870$ 
& $0.99974846$
\\

4
& $-0.5055 \pm 0.0014$
& $0.60508 \pm 0.00021$
& $0.687269$ 
& $0.99973803$
\\

5
& $-0.4938 \pm 0.0017$
& $0.5898 \pm 0.0003$
& $0.640092$ 
& $0.99977631$
\\

\hline
\end{tabular}

\caption{Fit parameters for the ansatz in Eq. \eqref{odd_Ninf_ansatz}. The reported uncertainties are conditional least-squares errors evaluated at fixed $\tilde{p}_n$. The uncertainties in $\tilde{p}_n$ were not estimated.}
\label{tab:odd_l_TL_fit}
\end{table}

\begin{figure}
	\centering
	\includegraphics[trim=1cm 2cm 0cm 0cm, clip, scale=0.34]{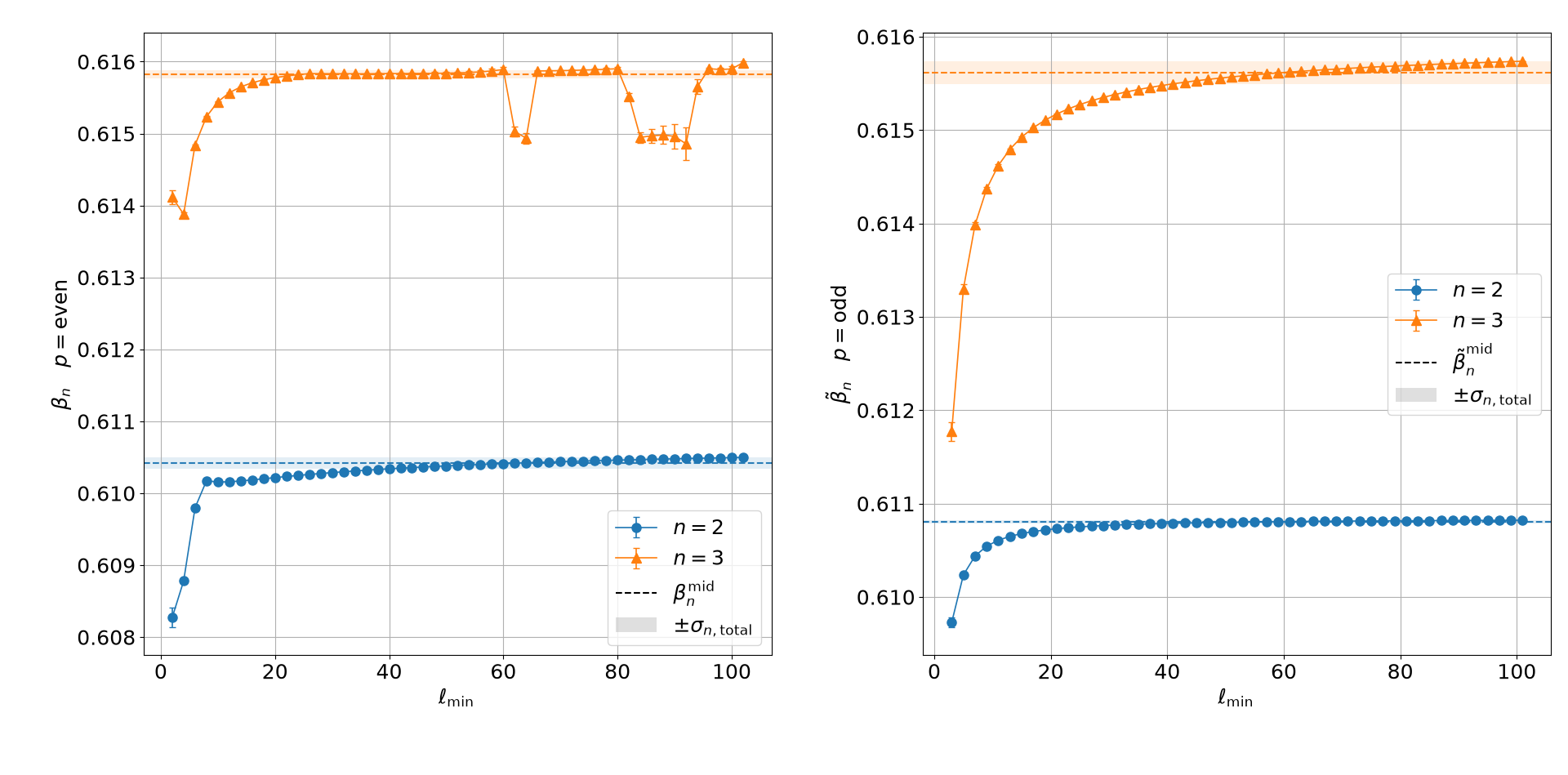}
	\caption{Dependence of the extrapolated constants on the lower fitting cutoff $\ell_{\min}$ for even $\ell$ (left) and odd $\ell$ (right), with $\ell_{\max}$ fixed at its largest available value. Dashed lines denote the midrange estimates $\beta_n^{\rm mid}$ over the selected stability windows, and shaded regions show $\beta_n^{\rm mid}\pm\sigma_{n,\rm total}$. The weak cutoff dependence after removal of the smallest subsystem sizes supports a stable nonzero thermodynamic-limit value.}
	\label{Plot_betaStability}
\end{figure}

We assess the sensitivity of the TL extrapolation to short-distance data by repeating the fits for $n=2,3$ while varying $\ell_{\min}$ and keeping $\ell_{\max}$ fixed. For each parity sector, we select a stable interval $\ell_{\min}\in[L_1,L_2]$ after the initial short-distance drift and define 
\begin{equation}
 \beta_n^{\rm mid}
 =
 \frac{\beta_n^{\max}+\beta_n^{\min}}{2},
 \qquad
 \sigma_{n,\rm total}
 =
 \sqrt{
 \left(
 \frac{\beta_n^{\max}-\beta_n^{\min}}{2}
 \right)^2
 +
 \sigma_{n,\rm LS}^2
 },
\end{equation}
\noindent where $\beta_n^{\max,\min}$ are the extrema within the selected interval and $\sigma_{n,\rm LS}$ is the largest conditional least-squares error in the same interval. As shown in Figure \ref{Plot_betaStability}, the extrapolated constants remain close to $0.61$ after discarding the smallest subsystem sizes (see Table \ref{tab:deduced-beta-window}), supporting a finite nonzero TL value.

%
\begin{table}
\centering
\small
\begin{tabular}{c c c}
\hline
Constant
& Plateau $[L_1,L_2]$
& Plateau value
\\
\hline

$\beta_2$
& $[40,102]$
& $0.61042\pm0.00008$
\\

$\beta_3$
& $[20,58]$
& $0.61582\pm0.00006$
\\

$\widetilde{\beta}_2$
& $[41,101]$
& $0.610805\pm0.000016$
\\

$\widetilde{\beta}_3$
& $[41,101]$
& $0.61562\pm0.00012$
\\

\hline
\end{tabular}%
\caption{Plateau values of $\beta_n$ (even) and $\widetilde{\beta}_n$ (odd). The final uncertainty is $\sigma_{n,\mathrm{total}}$.}
\label{tab:deduced-beta-window}
\end{table}

\newpage

\begin{figure}
    \centering
	\subfloat[\textit{Twist} for $n=2$.\label{Plot_fN_Asyms_data_M_comparison__n2}]{%
	\includegraphics[trim=0cm 0cm 0cm 0cm, clip, scale=0.35]{imgs_fN__Asym_M/PlotData__fNGS_Asym_n_l_N__M_comparison_n2_inset.png}
    }
    \hfill
    \subfloat[\textit{Twist} for $n=3$.\label{Plot_fN_Asyms_data_M_comparison__n3}]{%
	\includegraphics[trim=0cm 0cm 0cm 0cm, clip, scale=0.35]{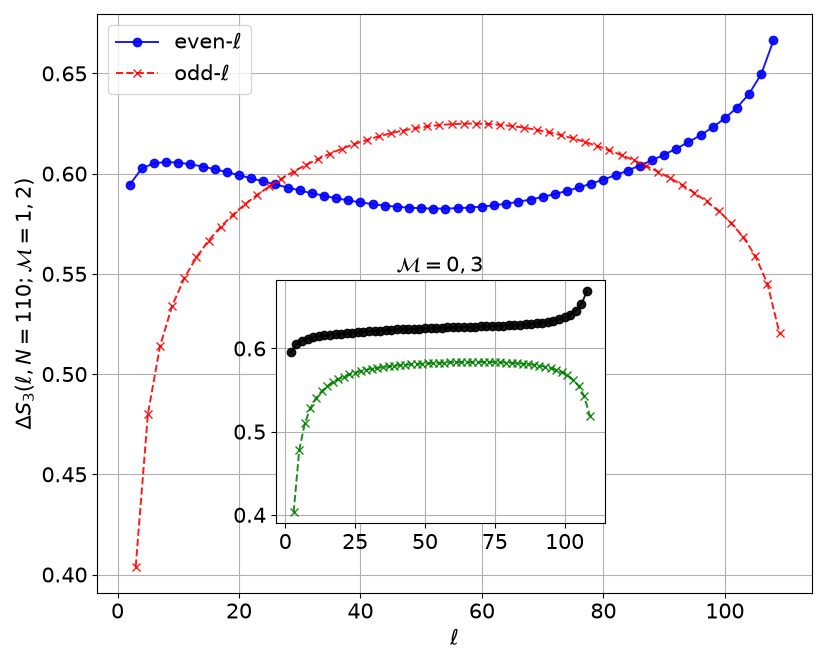}
    }

	\caption{Axial EA of the GS for $N=110$ and $n=2$ and $3$. For $\mathcal{M}=1,2$, the even- and odd-$\ell$ branches cross at $\ell/N\simeq1/4$ and $3/4$. No such crossings occur for $\mathcal{M}=0,3$ (inset).}
    \label{two_panels__M_twist}
\end{figure}

For even N, the GS subspace is fourfold degenerate. The $\mathcal{M}=1,2$ states exhibit a characteristic \textit{twist} relative to $\mathcal{M}=0,3$, as illustrated in the main text. This behavior persists for all accessible Rényi indices; Figure \ref{two_panels__M_twist} shows $n=2,3$. Since $\QAx\GSM{0,3}=0$ whereas $\QAx\GSM{1,2}=\mp \GSM{1,2}$, we interpret the twist as a local remnant of the global \textit{axial parity} $(-1)^{\QAx}$.\newline

Focusing on the $\mathcal{M}=0$ state, we analyzed the finite-size corrections 
\begin{equation}
R_n(\ell,N) := \Delta S_n(\ell,N) - \Delta S_n(\ell,\infty)
\label{remainder_definition}
\end{equation}

\begin{figure}
	\centering
	\includegraphics[trim=0cm 0cm 0cm 0cm, clip, scale=0.5]{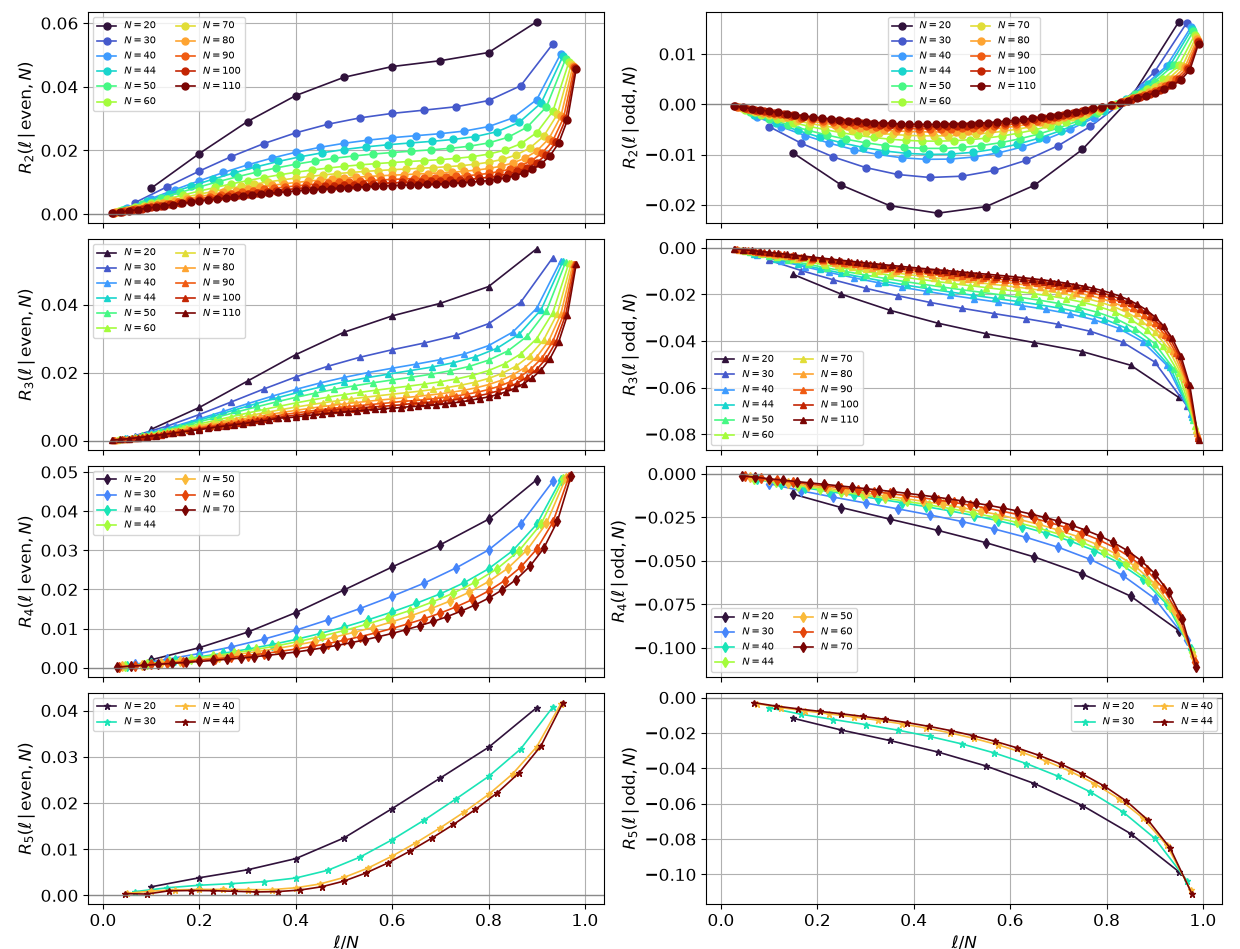}
	\caption{Finite-size corrections $R_n(\ell,N)$ for $\GSM{0}$. The even-$\ell$ (left) and 	odd-$\ell$ (right) sectors exhibit distinct scaling behavior for all Rényi indices considered.}
	\label{Plot_fN_R_Asyms_data}
\end{figure}

\begin{figure}
    \centering
    \subfloat[Even parity case.\label{Fit_fN_R_Asyms_data_even_l}]{%
	\includegraphics[trim=0cm 0cm 0cm 0cm, clip, scale=0.3]{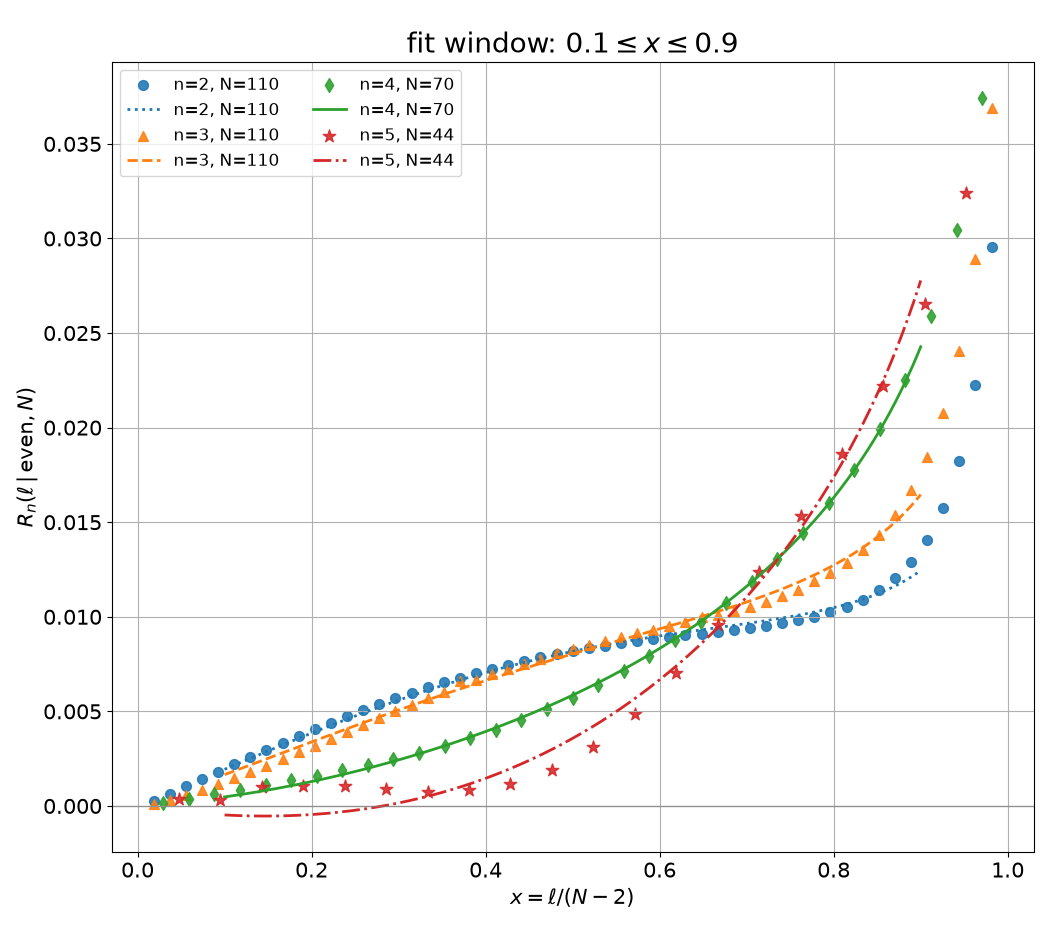}
    }
    \hfill
    \subfloat[Odd parity case.\label{Fit_fN_R_Asyms_data_odd_l}]{%
	\includegraphics[trim=0cm 0cm 0cm 0cm, clip, scale=0.3]{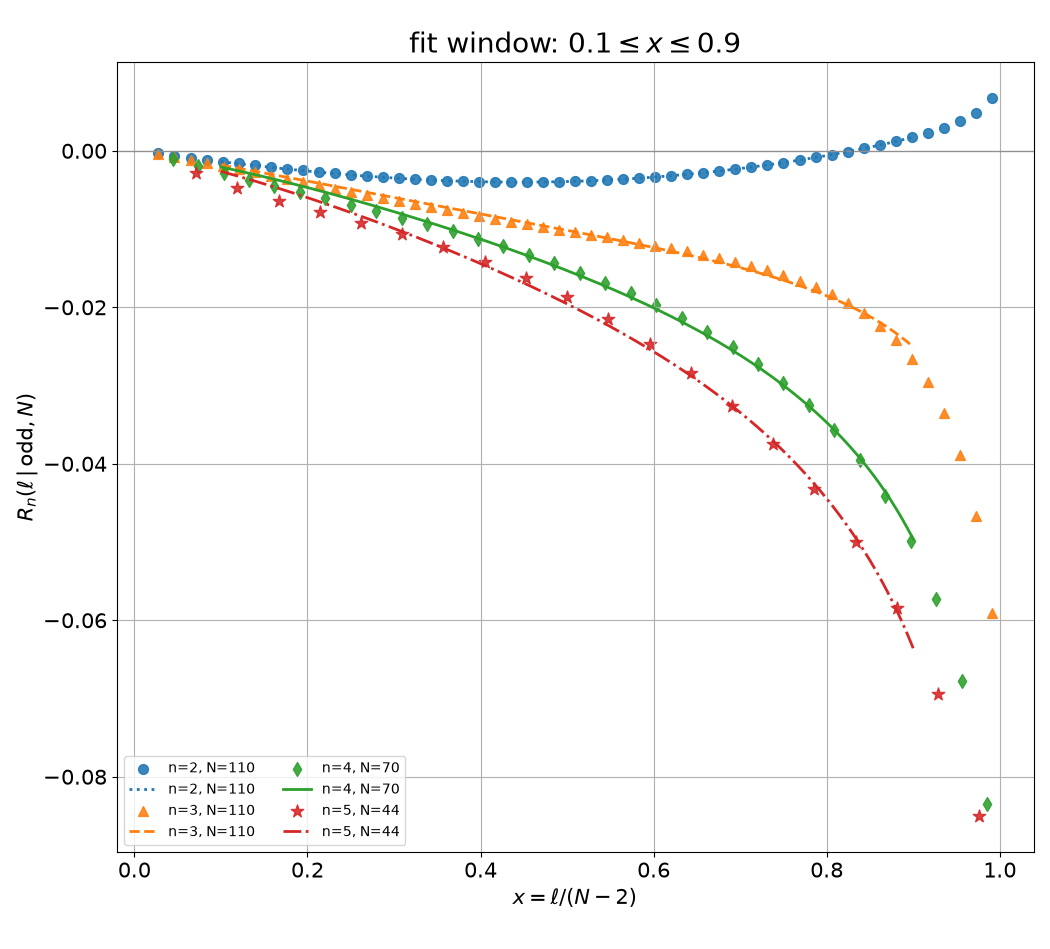}
    }

	\caption{Fits of $R_n(\ell,N)$ for $\GSM{0}$. The data window used for the fits is $x\in[0.1,0.9]$.}
    \label{two_panels_Fit_R_n_l_N}
\end{figure}

\begin{figure}
    \centering
    \subfloat[Even parity case.\label{Fit_fN_R_Asyms_data_even_l_scaling}]{%
	\includegraphics[trim=0cm 0cm 0cm 0cm, clip, scale=0.35]{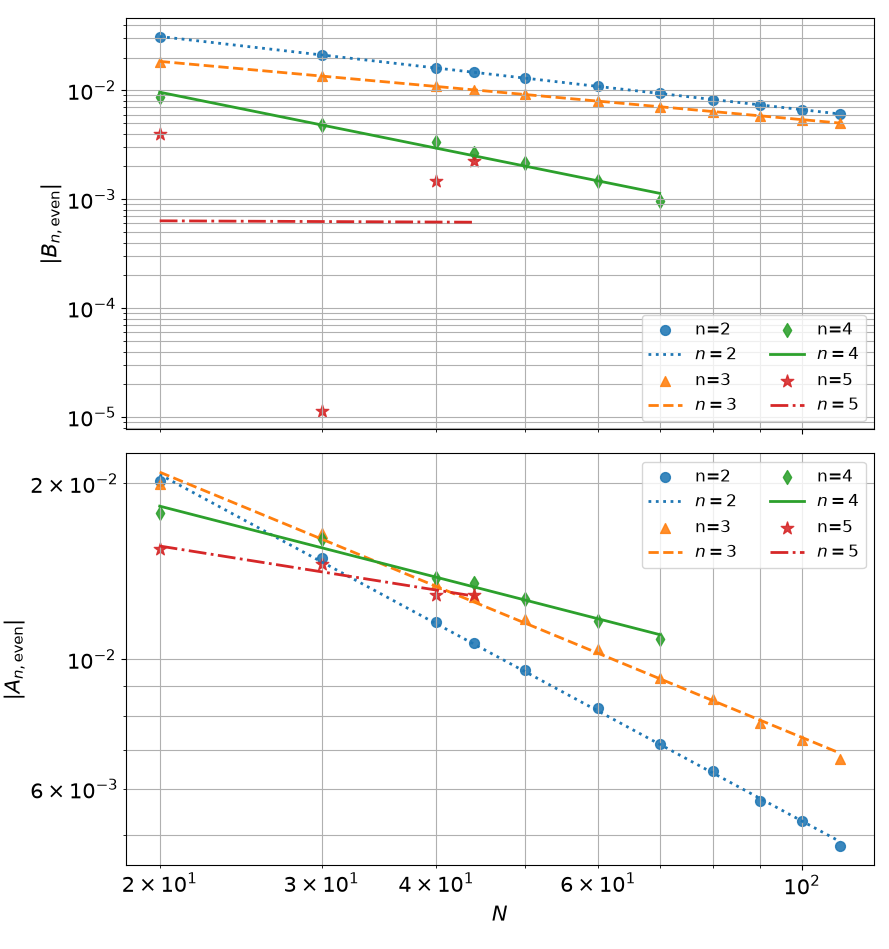}
    }
    \hfill
    \subfloat[Odd parity case.\label{Fit_fN_R_Asyms_data_odd_l_scaling}]{%
	\includegraphics[trim=0cm 0cm 0cm 0cm, clip, scale=0.35]{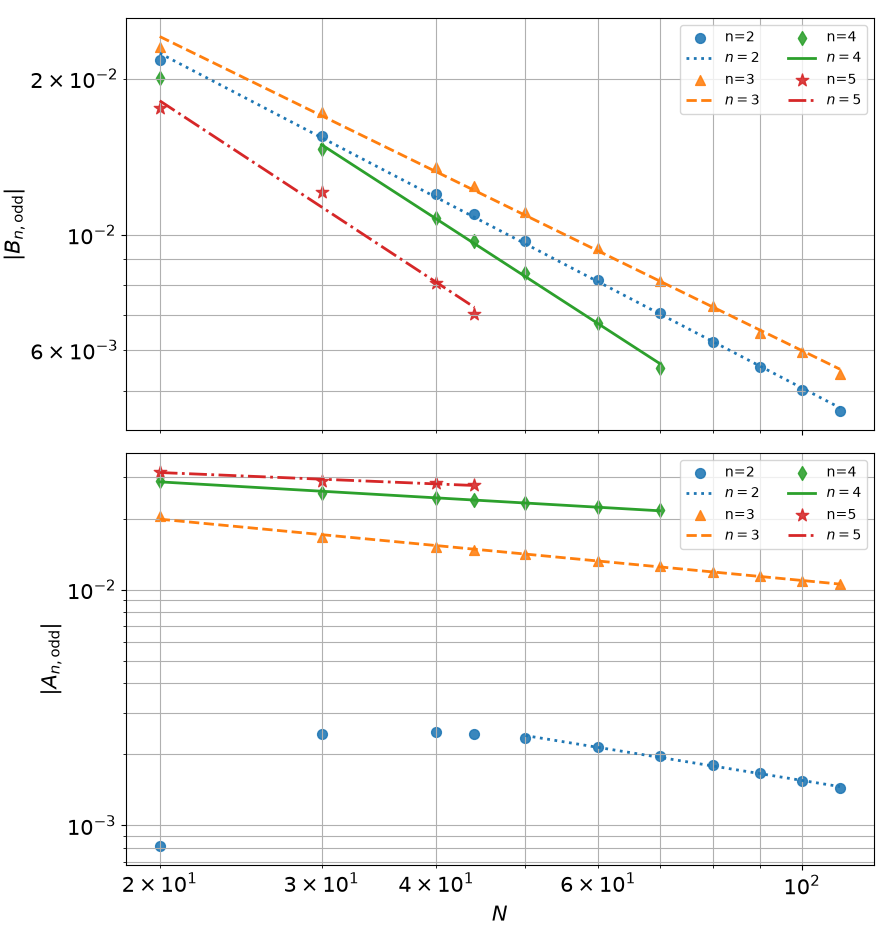}
    }

	\caption{The amplitudes $A_{n,\mathrm{p}}(N)$ and $B_{n,\mathrm{p}}(N)$ are fitted to algebraic decays in $N$. For $n=5$, the even-sector coefficient $B_{5,\mathrm{even}}(N)$ deviates from the power-law behavior observed for the other Rényi indices, likely because the accessible system sizes are too small. In the odd-$\ell$ sector at $n=2$, the coefficient $A_{2,\mathrm{odd}}(N)$ follows a power law only for $N\geq50$.}
    \label{two_panels_Fit_R_n_l_N__AmplitudeScaling}
\end{figure}

\noindent The raw data for $R_n(\ell,N)$ are shown in Figure \ref{Plot_fN_R_Asyms_data} for several system sizes $N$. After testing different scaling forms, we concluded that the corrections are well described by
\begin{equation}
R_n(\ell | \text{p} ,N) = B_{n,\text{p}}(N) \sin{(\pi x)} - A_{n,\text{p}}(N) \log{ \frac{\sin{(\pi x)}}{\pi x}}
\label{finite_size_corrections_ansatz}
\end{equation}
\noindent where $x=\ell/(N-2)$. A comparison between the data (for the largest $N$ available for each value of $n$) and this ansatz is shown in Figure \ref{two_panels_Fit_R_n_l_N}. Moreover, the fitting amplitudes $A_{n,\text{p}}(N)$, $B_{n,\text{p}}(N)$ exhibit finite-size algebraic decay $\propto 1/N^{q}$. The exponent $q$ depends on the coefficient ($A$ or $B$), the parity and Rényi index $n$: $q=q^{B/A}_{n,\text{p}}$ (see Figure \ref{two_panels_Fit_R_n_l_N__AmplitudeScaling}). We extract their values from weighted linear regressions 
\begin{equation}
\log{\left|C_{n,\text{p}}(N)\right|} = \alpha_{n,\text{p}}^{C} - q_{n,\text{p}}^{C} \log(N), \qquad C\in\{A,B\}
\end{equation}
\noindent with weights chosen as 
\begin{equation}
\omega_{n,\text{p}}^{C}(N) = \left( \frac{\left|C_{n,\text{p}}(N)\right|}{\sigma_{C_{n,\text{p}}(N)}} \right)^2
\end{equation}
\noindent corresponding to inverse-variance weighting after propagating the uncertainty $\sigma_{C_{n,\text{p}}(N)}$ from the fit of the $R_n(\ell,N)$ data to Eq.~\eqref{finite_size_corrections_ansatz}. The numerical results are reported in Table \ref{tab:q-exponents}.

\begin{table}
\centering
\begin{tabular}{c  c  c  c  c}
\hline
$n$
& $q^{B}_{n,\mathrm{even}}$
& $q^{A}_{n,\mathrm{even}}$
& $q^{B}_{n,\mathrm{odd}}$
& $q^{A}_{n,\mathrm{odd}}$
\\
\hline

2
& $0.980 \pm 0.006$
& $0.865 \pm 0.008$
& $0.9567 \pm 0.0018$
& $(0.680 \pm 0.007)^{*}$
\\

3
& $0.770 \pm 0.020$
& $0.688 \pm 0.014$
& $0.883 \pm 0.008$
& $0.355 \pm 0.005$
\\

4
& $1.89 \pm 0.07$
& $0.459 \pm 0.013$
& $1.14 \pm 0.04$
& $0.22 \pm 0.01$
\\

5
& $\dagger$
& $0.27 \pm 0.06$
& $1.14 \pm 0.07$
& $0.151 \pm 0.027$
\\

\hline
\end{tabular}

	\caption{Scaling exponents $q^{B}_{n,\text{p}}$ and $q^{A}_{n,\text{p}}$ for even and odd subsystem-length parities. The $\dagger$ denotes that the coefficient $B_{5,\mathrm{even}}(N)$ deviates from the power-law behavior in the available $N$ data. The $\text{}^*$ indicates that $A_{2,\mathrm{odd}}(N)$ follows a power law only for $N\geq50$.}
\label{tab:q-exponents}
\end{table}

\end{document}